\documentclass[12pt]{article}

\def\dalemb#1#2{{\vbox{\hrule height .#2pt
        \hbox{\vrule width.#2pt height#1pt \kern#1pt
                \vrule width.#2pt}
        \hrule height.#2pt}}}

\let\a=\alpha \let\b=\beta \let\g=\gamma \let\d=\delta \let\e=\epsilon
\let\z=\zeta  \let\th=\theta  \let\k=\kappa
\let\l=\lambda \let\m=\mu  \let\x=\xi  %\let\r=\rho
\let\s=\sigma \let\t=\tau    
 
      \let\G=\Gamma  \let\Th=\Theta \let\L=\Lambda
\let\X=\Xi  \let\S=\Sigma  \let\Y=\Psi
 
\let\la=\label  
  
\def\nn{\nonumber} \def\bd{\begin{document}} \def\ed{\end{document}}
\def\ds{\documentstyle} \let\fr=\frac \let\bl=\bigl \let\br=\bigr
\let\Br=\Bigr \let\Bl=\Bigl
\let\bm=\bibitem
\let\na=\nabla
\def\tU{{\widetilde U}}
\let\pa=\partial \let\ov=\overline
\def\ie{{\it i.e.\ }}
\newcommand{\be}{\begin{equation}}
\newcommand{\ee}{\end{equation}}
\def\ba{\begin{array}}
\def\ea{\end{array}}
\def\ft#1#2{{\textstyle{{\scriptstyle #1}\over {\scriptstyle #2}}}}
\def\fft#1#2{{#1 \over #2}}
\def\F#1#2{{ F_{#1}^{(#2)} }}
\def\cF#1#2{{ {\cal F}_{#1}^{(#2)} }}

\def\R{{\bf R}}
\def\sst#1{{\scriptscriptstyle #1}}
\def\oneone{\rlap 1\mkern4mu{\rm l}}
\def\e7{E_{7(+7)}}
\def\td{\tilde}
\def\wtd{\widetilde}
\def\im{{\rm i}}
\def\bog{Bogomol'nyi\ }
\newcommand{\ho}[1]{$\, ^{#1}$}
\newcommand{\hoch}[1]{$\, ^{#1}$}
\newcommand{\bea}{\begin{eqnarray}}
\newcommand{\eea}{\end{eqnarray}}
\newcommand{\ra}{\rightarrow}
\newcommand{\lra}{\longrightarrow}
\newcommand{\Lra}{\Leftrightarrow}
\newcommand{\ap}{\alpha^\prime}
\newcommand{\bp}{\tilde \beta^\prime}
\newcommand{\cB}{{\cal B}}
\newcommand{\cO}{{\cal O}}
\newcommand{\vecx}{\vec{x}}
\newcommand{\vecy}{\vec{y}}
\newcommand{\vecp}{\vec{p}}
\newcommand{\vecq}{\vec{q}}
\newcommand{\tr}{{\rm tr} }
\newcommand{\Tr}{{\rm Tr} }
\newcommand{\NP}{Nucl. Phys. }

\newcommand{\cL}{{\cal L}}
\newcommand{\cA}{{\cal A}}
\newcommand{\cD}{{\cal D}}

\def\sst#1{{\scriptscriptstyle #1}}
\def\0{{\sst{(0)}}}
\def\1{{\sst{(1)}}}
\def\2{{\sst{(2)}}}
\def\3{{\sst{(3)}}}
\def\4{{\sst{(4)}}}
\def\5{{\sst{(5)}}}
\def\6{{\sst{(6)}}}
\def\7{{\sst{(7)}}}
\def\8{{\sst{(8)}}}
\def\ve{\varepsilon}
\def\vf{\varphi}
\def\F{\Phi}
\def\wg{\wedge}

\newcommand{\tamphys}{\it %Center for Theoretical Physics,\\
}

\newcommand{\auth}{AUTHORS}

\def\thb{\bar{\theta}}
\def\Thb{\bar{\Theta}}
\def\barp{\bar{p}}
\def\barq{\bar{q}}
\def\barc{\bar{c}}
\def\bard{\bar{d}}
\def\e{\epsilon}

\def \bi{\bibitem}
\def \la {\label}

\def \l {\lambda}
\def\foot{\footnote}
\def \tl  {{\tilde \l}}
\def \sql {{\sqrt \l}}
\def \adss {$AdS_5 \times S^5$\ }
\newcommand{\rf}[1]{(\ref{#1})}
\def \ov {\over}

\def\th{\theta}
\def\Th{\Theta}
\def\vth{\vartheta}
\def\btheta{{\bar\theta}}
\def\ttheta{{{\tilde\theta}}}
\def\bttheta{{{\bar\ttheta}}}
\def\vth{\vartheta}

\def\ra{\rightarrow}
\def\N{{\cal N}}
\def\F{{\cal F}}
\def\uM{\underline{M}}
\def\uN{\underline{N}}
\def\uP{\underline{P}}
\def\cc{\circ}
\def\eqv{\equiv}

\def\ni{\noindent}

\def\Ep{E^{{}^{(+)}}}
\def\Em{E^{{}^{(-)}}}

\def\Mp{M^{{}^{(+)}}}
\def\Mm{M^{{}^{(-)}}}

\def \ha{{1\ov 2}}

\def\r{\rho}

\def\Y{{\rm Y}}
\def\X{{\rm X}}
\def\tY{\tilde{\rm Y}}
\def\tX{\tilde{\rm X}}
\def\dY{\dot{\rm Y}}
\def\dX{\dot{\rm X}}

\def \J {\mathcal{J}}
\def \del {\partial}

\def\dF{\dot{F}}
\def\dG{\dot{G}}
\def\df{\dot{f}}
\def \E {{\cal E}}
\def \S {{\cal S}}
\def \J {{\cal J}}

\def\ms{\mathcal{S}}
\def\mj{\mathcal{J}}
\def\soj{\fr{\ms}{\mj}}
\def \R {{\bf R}}
\def \om {\omega}
\def \bE {\bar E}
\def \x {{\cal X}}

\def \bi{\bibitem}
\def \la {\label}

\def \l {\lambda}
\def\foot{\footnote}
\def \tl  {{\tilde \l}}
\def \sql {{\sqrt \l}}
\def \adss {$AdS_5 \times S^5$\ }
\def \ov {\over}

\def \varpi {{\rm w}}

\def\thb{\bar{\theta}}
\def\Thb{\bar{\Theta}}
\def\zb{\bar{z}}
\def\psib{\bar{\psi}}
\def\barp{\bar{p}}
\def\barq{\bar{q}}
\def\barc{\bar{c}}
\def\bard{\bar{d}}
\def\e{\epsilon}

\def\At{\tilde{A}}
\def\Bt{\tilde{B}}
\def\Ct{\tilde{C}}
\def\Dt{\tilde{D}}
\def\Et{\tilde{E}}
\def\Ft{\tilde{F}}
\def\Gt{\tilde{G}}
\def\Mt{\tilde{M}}
\def\at{\tilde{a}}
\def\bt{\tilde{b}}
\def\ct{\tilde{c}}
\def\dt{\tilde{d}}
\def\et{\tilde{e}}
\def\ft{\tilde{f}}
\def\gt{\tilde{g}}
\def\ola{\overleftarrow}
\def\ora{\overrightarrow}
\def\at{\tilde{\a}}

\def\ps{\rlap{\, /}\;\,p }
\def\ks{\rlap{\, /}\;\,k }

\def\gym{g_{YM}}

\def\adot{\dot{a}}
\def\bdot{\dot{b}}

\newcommand{\PL}{{\em Phys.\ Lett.\ }}
\newcommand{\PR}{{\em Phys.\ Rev.\ }}
\newcommand{\PRP}{{\em Phys.\ Rep.\ }}
\newcommand{\CMP}{{\em Comm.\ Math.\ Phys.\ }}
\newcommand{\MPL}{{\em Mod.\ Phys.\ Lett.\ }}
\newcommand{\PRL}{{\em Phys.\ Rev.\ Lett.\ }}
\newcommand{\IJMP}{{\em Int.\ J.\ Mod.\ Phys.\ }}

\begin{document}
\overfullrule=0pt
\parskip=2pt
\parindent=12pt
\headheight=0in \headsep=0in \topmargin=0in
\oddsidemargin=0in

\vspace{ -3cm}
\thispagestyle{empty}
%\vspace{1cm}
%\begin{flushright}
%Preprint DFPD 01/TH/\\
%hep-th/
%\end{flushright}

\begin{center}

{\Large\bf Open string engineering of D-brane geometry
 %\\
 %\vspace{0.1cm}
  }

 \vspace{.5cm} { I.Y. Park\footnote{visiting KIAS for summer}  }\\
 \vskip 0.2cm

\vspace{0.5cm}
{\it Korea Institute for Advanced Study\\
Seoul 130-012, Korea \\
}
 \vspace{0.3cm}
 and\\
 \vspace{0.3cm}
{\it Department of Chemistry and Physics, University of Arkansas at Pine Bluff\\
Pine Bluff, AR 71601, USA \\
inyongpark05@gmail.com}

%\vspace{0.8cm}
%{\small\it Department of Physics, The Ohio State University\\
%Columbus, OH 43210, USA }

\end{center}

 \vspace{0.1cm}

 \begin{abstract}
 %%%%%%%%%%%%%%%%%%%%%%%%%%%%%%%%%
\ni One-loop scattering on a stack of D3 branes was considered in
arXiv:0801.0218 [hep-th]. Divergence was found and its cancelation
mechanism was proposed, wherein it was conjectured that the D-brane
geometry be introduced in the form of counter vertex operators. Here
we verify the conjecture at the first few leading orders in an
expansion method that we call large-$r_0$ expansion. We comment on
the relation with the Fischler-Susskind mechanism and discuss the
implications of our result for AdS/CFT.

\end{abstract}
\newpage

\setcounter{equation}{0}
\setcounter{footnote}{0}
\setcounter{section}{0}

%\renewcommand{\theequation}{1.\arabic{equation}}
% \setcounter{equation}{0}

%%%%%%%%%%%%%%%%%%%%%%%%%%%%%%%%%%%%%%%%%%%%%%%%%%%%%%%%%%%%%%%%
\section{Introduction}
%%%%%%%%%%%%%%%%%%%%%%%%%%%%%%%%%%%%%%%%%%%%%%%%%%%%%%%%%%%%%%%%
A D-brane is a hyperplane where the end points of an open string can
be attached \cite{pol}. When two or more open strings come across
they will scatter each other. Studying the scattering will be
interesting and relevant for several reasons.\footnote{In the past
scattering on D-branes was studied in the NSR formulation by several
authors \cite{Hashimoto:1996bf}. We will use the Green-Schwarz (GS)
formulation which is almost inevitable for the things that we try to
do.} For example, knowledge on scattering may shed some light on the
better understanding of AdS/CFT and its derivation, which has been
our main motivation. The derivation in turn may provide a new
paradigm for the unification of gauge theory and gravity. When a
 phenomenologically more realistic models of a D-brane configuration
becomes available it may also be necessary to consider scattering of
states not only at the low energy field theory level (which may or
may not be renormalizable) but also at the full level of open
string.

For the actual study one must first construct vertex operators for
the external scattering states. This has been carried out in the GS
formulation in one of our previous works taking the D3 brane case as
an example \cite{Park:2007mc}. On a D3 brane the D9-brane multiplet
gets resolved into two multiplets which we call the scalar multiplet
and the vector multiplet in analogy with the N=2 susy field theory.
Subsequently various tree and one-loop amplitudes were computed
\cite{Park:2008sg}. One loop divergence structure was obtained. It
was noted that the divergence structure does not share the nice
feature of the D9 brane, which seems to suggest that it may require
a more radical measure to remove. The deviation from the D9 brane is
due to the different structure of the zero modes. A proposal for
cancelation of the divergence was put forward in \cite{Park:2008sg}:
it should be possible to absorb the divergence by adding ``counter
vertex operators" of composite nature. They are to be constructed
out of the open string fields.

 It was conjectured that the precise forms of the
vertex operators will have a link to the geometry induced by the
D-branes.\footnote{A direct connection between the quantum effects
and geometry is not entirely new. For example the map obtained in
\cite{Gonzalez-Rey:1998uh} can be interpreted in this context. The
link between divergence and geometry goes back to the
Fischler-Susskind mechanism \cite{Fischler:1986ci,Fischler:1986tb}.
We comment on the relation in the conclusion.} According to the
conjecture the geometry should provide a guidance to find the
counter vertex operators. It would be a hard task to construct them
without such aid. Once fixing the form of the counter vertex
operator the geometry should be taken as an out-come: it arises {\em
as a result of the flat space analysis}. It is a secondary
by-product of open string loop effects, hence the title of the
paper.  Nowhere in the construction the explicit closed string
degrees of freedom are used. The composite vertex operators might be
interpreted as representing a closed string state but that is,
together with the by-product geometry, as close as it gets to the
close string. The whole construction so far is based on the purely
open string frame work.

Some preliminary computation was presented in \cite{Park:2008sg} on
the amplitudes with the counter vertex operators inserted. In this
work we initiate a much more systematic verification of the
conjecture focusing on two cases, the four scalar amplitude and the
four vector amplitude. The amplitudes at tree and one loop have been
obtained previously without the insertion of the couner vertex
operators: here we compute at tree level
 %%%
 \bea
 <V_sV_sV_sV_s\,V_G> \mbox{and} <V_gV_gV_gV_g\,V_G>
 \eea
 %%%
 where $V_s$ ($V_g$) denotes the scalar vertex operator (the
 vector vertex operator) and $V_G$ the counter
vertex operator. The subscript ``G" represents its proposed origin,
the geometry. We interchangeably call the counter vertex operator
the geometry vertex operator.

The detailed construction of $V_G$ is presented in the appendix. The
basic idea is to start from the GS action in a generic curved
background. The action was constructed relatively recently in
\cite{Cvetic:1999zs}\cite{Sahakian:2004gy}\cite{Mizoguchi:2002qy}.
In place of each supergravity field one substitutes the supergravity
solution for the D3 brane geometry. For a perturbative analysis one
makes an expansion which we call large $r_0$-expansion wherein one
introduces coordinates, $X^m=X_0^m,\; r_0=\sum_m X_0^mX_0^m$, for a
location that is far away from the center branes. Then one makes an
expansion of the resulting non-linear sigma model action around that
point. Then one identifies the fields in the external scattering
states as the fluctuation fields in the shifted coordinates. Why are
such an expansion and identification necessary?  From a practical
viewpoint the large-$r_0$ expansion seems inevitable for the
perturbative computation. Put differently the connection between the
geometry and the loop effects are made where it can be made in the
brane geometry. Obviously it is not the near-horizon region that we
look into. None the less one can make a connection to AdS geometry
and AdS/CFT once one establishes the role of D-brane geometry. We
will have more remarks on this in the conclusion.

As stated above our intention is to exclusively use the open string
degrees of freedom. For one thing it would be more economical, at
least from the standpoint of unifying degrees of freedom, than using
the closed string degrees of freedom as well. However, as for the
geometry one may question whether it really is necessary. In other
words, wouldn't it be possible to cancel the divergence using a flat
space action since that would be much simpler if things can work
that way?
 The fact that the D9 brane way of canceling the
divergence (i.e., by shifting the string tension \cite{gsw}) does
not work for the D3 brane case can be seen as follows. As in the D9
case the D3 brane one-loop produces divergence with
 exactly the same tree level kinematic- and gauge- structure. Let's
 attempt to cancel the divergence by shifting the tension of the
 {\em flat} space action. If one considers the flat space action there
 is no distinction between the D3 and D9 (since only the bd
 conditions
 are different): the action is
%%%
 \bea
 S&=&-\fr12\int\; (T\pa_\a X^i \pa^\a X^i-\fr{i}{\pi}\bar{S}^a \r^\a\pa_\a
 S^a)
 \eea
 %%%
The counter vertex operator that results as a consequence of varying
the string tension is $\pa_\t X^i \pa\t X^i-\pa_\s X^i \pa\s X^i
=\pa_\t X^u \pa_\t X^u-\pa_\s X^m \pa_\s X^m$ where we have omitted
irrelevant factors. The missing terms have dropped due to the fact
that the vertex operators are considered at $\s=0$. Recall that the
one loop results are such that the scalar four point and the vector
four point amplitudes have the same signs. It implies that with the
given relative signs one can not cancel the divergence of the scalar
loop and the vector loop at the same time. It is to be contrasted
with how things go in the D3-brane background. There the additional
sign comes about, as we will see, due to the curved metric factors,
$H^{1/2}$ and $H^{-1/2}$, making the geometry vertex operator
$-\fr{q}{2} \pa_\t X^\m \pa\t X^\m-\fr{q}{2}\pa_\s X^m \pa\s X^m$ in
the leading order. It can be read off from the quadratic order
action in fermion,
%%%
 \bea
&&  \;-\fr{1}2\sqrt{h}\;h^{ij}
 \left(\pa_i X^u \pa_j X^v \eta_{uv} (H^{-1/2}-1)+
 \pa_i X^m \pa_j X^n \eta_{mn} (H^{1/2}-1)\right)
  \nn\\
 %%%%%%%%%%%%%%%%%%%%%%%%%%%%%%%%
 && -\fr{i}{p^+}
   (\sqrt{h}\;h^{ij}-\ve^{ij})\pa_i X^+ (H^{-1/4}-1)
         (S\pa_jS)
  \eea
  %%%
With the flip of the sign the counter vertex operator has a
potentially right form to work, and it does work as we will see
below. Note also that the curved space provides the needed factor of
the open string coupling constant, $g$, through $q$ (defined in
(\ref{r0q}) below) automatically. The sign contradiction alone is
sufficient to rule out the flat space action. But there are some
other ominous features that makes the flat action unlikely. For
example, since the string tension $T$ appears only in the bosonic
part the fermionic coordinates will not play a role what so ever in
any arbitrary order. Although the fermionic term does not seem to
play a role either in the examples that we consider in this
work\footnote{Remember that we are considering a first few leading
orders of particular cases.}, it simply can not be true in general.
Another unfavorable feature of the flat space counter vertex
operator is associated with the two loop structure. The final form
of a two-loop four point amplitude will have a single integration as
far as the world-sheet locations of the vertex operators are
concerned as with the corresponding one-loop amplitude. In attempt
to cancel the divergence one would have to insert two vertex
operators of the form $\sim g^4\pa X(y_1)\pa X(y_1) \pa X(y_2)\pa
X(y_2)$. Therefore even after performing one of the two
$y$-integration there is one too many integration compared with the
two-loop result: it seems unlikely that the flat space operator will
succeed. On the contrary the geometry after the large-$r_0$
expansion naturally produces a term at $q^2\sim g^4$-order at a
single $y$-location. Therefore, it seems that the flat space option
is ruled out. The real question is whether the D-brane geometry does
the job and, if not, what else.

The rest of the paper is organized as follows. In sec 2, we make a
summary of results and present some of the salient features of the
computations detailed in sec 3. We point out that the results are
verification, at leading orders, of the conjecture that was put
forward in \cite{Park:2008sg}.
  In section 3, we begin by putting together
several ingredients for the forthcoming computations through a brief
review. We quote the full  expression (i.e., expression prior to the
large $r_0$ expansion) of the geometry vertex operator that is
obtained in the appendix. After the large-$r_0$ expansion we carry
out the four point amplitude computation for the scalar multiplet
and the vector multiplet for the first two orders of the expansion.
Partial results are mentioned on the third leading order. Many parts
of the computations are prohibitively
 long for manual computation. Much of the computation has been Mathematica-coded.
In the conclusion we discuss various issues such as the implications
of our results for AdS/CFT (especially the stronger form thereof),
some of the loose points raised in the main body, future directions.
In the appendix we outline how to obtain the geometry vertex
operator.

%%%%%%%%%%%%%%%%%%%%%%%%%%%%%%%%%%%%%%%%%%%%%%%%%%%%%%%%%%%%%%%%
\section{Summary of results }
%%%%%%%%%%%%%%%%%%%%%%%%%%%%%%%%%%%%%%%%%%%%%%%%%%%%%%%%%%%%%%%%

The section that follows the present one contains lengthy and
tedious pieces of computations. It may be a good idea to have a
summary of the results before we embark on heavy computation. To
prove the conjecture, first we must show that the correlator
$<VVVV\;V_G>$ has precisely the same kinematic and momentum
structure as the corresponding one loop (and tree since there are
the same) result of $<VVVV>$. The computation below seems to suggest
a pattern on how this is achieved: a first few leading order terms
in $V_G$ alone produce the desired structure with the higher order
terms yielding vanishing contributions.

The result of the appendix, (\ref{startingaction5}), suggests the
following form of the counter vertex operator with $S$ being the
fermionic coordinate,
%%%
 \bea
\pi V_G
 =&&  \;-\!\!\fr{1}2\sqrt{h}\;h^{ij}
 \left(\pa_i X^u \pa_j X^v \eta_{uv} (H^{-1/2}-1)+
 \pa_i X^m \pa_j X^n \eta_{mn} (H^{1/2}-1)\right)
  \nn\\
 %%%%%%%%%%%%%%%%%%%%%%%%%%%%%%%%
 && +\fr1{2p^+} \{ \nn\\
  && -2i(\sqrt{h}\;h^{ij}-\ve^{ij})\pa_i X^+ (H^{-1/4}-1)
         (S\pa_jS)  \nn\\
 &&+\fr{i}4 (\sqrt{h}h^{ij}-\ve^{ij})\pa_i X^+ H^{-7/4}\fr{H'}{r} \pa_jX^u
        X^m\; (S\g^{um}S)\nn\\
  &&    -\fr{i}4 (\sqrt{h}h^{ij}-\ve^{ij})\pa_i X^+ H^{-5/4}\fr{H'}{r}
       \pa_jX^m X^n\;(S\g^{mn}S)\nn\\
        && \quad\quad\} \nn\\
 %%%%%%%%%%%%%%%%%%%%%%%%%%%%%%%%%%
       &&+\fr1{4(p^+)^2}\sqrt{h}h^{ij}\pa_i X^+\pa_j X^+\;H^{-1/2}\{\nn\\
  &&-\fr{17}{1536}\k_1(S\g^{uv} S)( S\g^{uv} S)+\left[
  \fr{43}{768}\k_1+\fr1{192}\k_2
  \right](S\g^{au} S)( S\g^{au} S) \nn\\
  &&-\left[
  \fr1{192}\k_2 +\fr1{128}\k_1\right] (S\g^{ab} S)( S\g^{ab} S)\nn\\
  &&+X^aX^b \fr{1}{r^2}\left[\fr{31}{768}\k_1-\fr1{32}\k_2
  \right](S\g^{au} S)( S\g^{bu} S)\nn\\
  &&+X^aX^b \fr{1}{r^2}\left[
  +\fr1{32}\k_2+\fr{29}{384}\k_1\right](S\g^{ac} S)( S\g^{bc} S) \nn\\
 &&\} \label{startingactionfinal}
 \eea
 %%%
 where
 %%%
 \bea
 \k_1=H^{-5/2}(H')^2,\quad\quad \k_2=H^{-3/2}H'\fr1r,\quad\quad
 H(X^m)=1+\fr{4\pi g^2  \a'^2}{r^4}
 \eea
 %%%
 In the right hand side of the third equation we have replaced the
 closed string coupling constant by the open string coupling
 constant.
For a perturbative approach we expand the operator around a point,
$X_0^m$ with $(X_0^m)^2\equiv r_0^2$, that is far away from the
center branes. Because of the SO(6) rotational symmetry of the brane
configuration the individual coordinate $X_0^m$ will only appear
through $r_0$ which we will fix later. To illustrate the large $r_0$
expansion consider the function $H$. Define
%%%
 \bea
 r_0^4 =\L^4\, \a'^2,\quad
 q=\fr{4\pi g^2}{\L^4} \label{r0q}
 \eea
 %%%
 where $\L$ is a dimensionless parameter that measures the norm of
 $r_0$ in terms of $\sqrt{\a'}$.
 Shifting $X^m \ra
X^m+ X_0^m $ one gets
%%%
 \bea
H(X+X_0)=&& 1 + {q} - \frac{4\,q\,{X_0}\cdot X}{{ {r_0}}^2} +
  q\,\left( \frac{-2\,r^2}{{{r_0}}^2} +
     \frac{12\,{\left( {X_0}\cdot X \right) }^2}{{{r_0}}^4}
     \right)+\cdots
 \eea
 %%%
It is nice to note that due to the dimensional regularization only a
finite number of terms contribute for a fixed number of external
states and a fixed space-time loop order. For example in the case of
four point scattering we should expand up to (and including)
$X^4$-order: higher order terms do not make
contributions.\footnote{For the purely vector multiplet scattering
it is even simpler since a longitudinal coordinate, $X^u$, and a
transverse coordinate, $X^m$, do not contract each other: one can
simply set $X^m=X_0^m$. Incidentally, this does not make the vector
case simpler. The reason is that the external states come with
$e^{ikX}$-factor which contains the longitudinal coordinates. For
the $q$-order four point amplitudes that we consider only the
quadratic terms contribute. } The expansion parameters are taken as
$\fr{1}{r_0}$ (or $\L$) and $q$. Since we are dealing with the
one-loop divergence, only the linear terms in $q$ may be kept. It
seems that in the leading order of $\fr{1}{r_0}$ all of the
S-quartic terms drop basically because of the fermionic equation of
motion.

%%%%%%%%%%%%%%%%%%%%%%%%%%%%%%%%%%%%%%%%%%%%%%%%%%%%%%%%%%%%%%%%
\subsection{scalar multiplet scattering}
%%%%%%%%%%%%%%%%%%%%%%%%%%%%%%%%%%%%%%%%%%%%%%%%%%%%%%%%%%%%%%%%

The kinematic structure of the one-loop divergence
\cite{gsw}\cite{Park:2008sg} is
 %%%
 \bea
 <V_sV_sV_sV_s >\sim \fr1{\e}
 \fr14(su\;\xi_1\cdot \xi_4\;\xi_2\cdot \xi_3
 +tu\;\xi_1\cdot \xi_2\;\xi_3\cdot \xi_4
 +st\;\xi_2\cdot \xi_4\;\xi_1\cdot \xi_3)
 \eea
 %%%
 where $\e$ is a infinitesimal parameter.
 What we want to show, therefore, is
 %%%
 \bea
 <V_sV_sV_sV_s\;V_{G} >\sim
 \fr14(su\;\xi_1\cdot \xi_4\;\xi_2\cdot \xi_3
 +tu\;\xi_1\cdot \xi_2\;\xi_3\cdot \xi_4
 +st\;\xi_2\cdot \xi_4\;\xi_1\cdot \xi_3)
 \eea
 %%%
Here and below we have suppressed a common factor
$\fr{\G(-\a's)\G(-\a't)}{\G(1-\a's-\a't)}$. It is a necessary
condition. Making it sufficient will give a relation between $\e,\L$
and $\e_y$ as we will discuss towards the end of sec 2. We define
$\e_y$ below. We break $V_G$ into the power series expansion in
$\fr1{r_0}$,
 %%%
 \bea
 V_G=V_{G,r_0^{-4}}+V_{G,r_0^{-5}}+V_{G,r_0^{-6}}+\cdots
 \eea
 %%%
As indicated the leading order vertex operator comes with $\fr1{r_0^4}$.
We will work out the explicit form below and show that
%%%
 \bea
\pi V_{G,r_0^{-4}}
 && = \fr{q}4  \;
 \left(-\pa_\s X^m \pa_\s X^m-\pa_\t X^u \pa_\t X^u
  +il^2\,(- S\pa_\t S- S\pa_\s S )\right) \label{leadingV2quoted}
  \eea
  %%%
 With this one gets
 %%%
 \bea
 <V_sV_sV_sV_s\;V_{G,r_0^{-4}} >=\fr{4\pi g^2}{\e_y \L^4}
 \fr14(su\;\xi_1\cdot \xi_4\;\xi_2\cdot \xi_3
 +tu\;\xi_1\cdot \xi_2\;\xi_3\cdot \xi_4
 +st\;\xi_2\cdot \xi_4\;\xi_1\cdot \xi_3)
 \eea
 %%%
Other than the factor in front $\fr{4\pi g^2}{\e_y \L^4}$ it is
precisely the kinematic factor of the corresponding tree (and the
one-loop) diagram. The parameter $\e_y$ is infinitesimal and
introduced to regulate the divergence of the amplitude with the
geometry vertex operator inserted. One sees that by adjusting
$\fr{1}{\e_y \L^4}$ one can absorb the one-loop divergence. One of
the nice things about the result is that the computation does not
produce any finite part: the only power of $\e_y$ that appears is
$\fr{1}{\e_y}$. As a matter of fact in all the computations that we
have performed so far it remains true. The next leading order vertex
operator is
%%%
 \bea
 \pi V_{G,r_0^{-5}}=&&    -\fr{i}4 (\sqrt{h}h^{ij}-\ve^{ij})\pa_i X^+ H_0^{-5/4}\fr{H_0'}{r_0}
       \pa_jX^m X_0^n\;(S\g^{mn}S)\nn\\
 &&+\fr{i}4 (\sqrt{h}h^{ij}-\ve^{ij})\pa_i X^+ H_0^{-7/4}\fr{H_0'}{r_0} \pa_jX^u
        X_0^n\; (S\g^{un}S)
 \eea
 %%%
At this order the amplitude turn out to vanish,
 %%%
 \bea
 <V_sV_sV_sV_s\;V_{G,r_0^{-5}} >=0
 \eea
 %%%
In the third leading order the geometry vertex operator is
 %%%
 \bea
V_{G,r_0^{-6}}= &&-\fr{i\pa_i X^+}{2p^+}
   (\sqrt{h}\;h^{ij}-\ve^{ij})
  \left[X^n X^n\;S\pa_jS+\pa_jX^u X^n\; (S\g^{un}S))\right.\nn\\
  &&\left.-\pa_jX^m X^n\;(S\g^{mn}S) \right]  \nn\\
 %%%%%%%%%%%%%%%%%%%%%%%%%%%%%%%%%%
       &&-\fr1{192}\sqrt{h}h^{ij}\fr{\pa_i X^+\pa_j X^+}{(p^+)^2}\;
       \{(S\g^{au} S)( S\g^{au} S) -  (S\g^{ab} S)( S\g^{ab} S)\}
  \eea
  %%%
With increasing number of the fields the computation becomes quickly
complicated even for the machine computing. Although we have not
entirely completed computation we have carried out some of the
correlators. For example we have checked that
 %%%
 \bea
 <XXXX\;V_{G,r_0^{-6}}>=0
 \eea
 %%%
There are several other corrrelators that we have checked. Based on
the computations so far we expect that the correlator at this order
will vanish, $<V_sV_sV_sV_s\;V_{G,r_0^{-6}}>=0$. We mention the
reason for the expectation in the conclusion.

%%%%%%%%%%%%%%%%%%%%%%%%%%%%%%%%%%%%%%%%%%%%%%%%%%%%%%%%%%%%%%%%
\subsection{vector multiplet scattering}
%%%%%%%%%%%%%%%%%%%%%%%%%%%%%%%%%%%%%%%%%%%%%%%%%%%%%%%%%%%%%%%%

A similar pattern is found in the case of the vector scattering: the
leading order terms in $V_G$ produces the desired kinematic
structure and the higher order terms yield vanishing results. Recall
the kinematic structure of the tree level scattering without $V_G$,
%%%
 \bea
 K=&&-\fr14(st\;\z_1\cdot \z_3\;\z_2\cdot \z_4+
     su\;\z_2\cdot \z_3\;\z_1\cdot \z_4
     +tu\;\z_1\cdot \z_2\;\z_3\cdot \z_4)\nn\\
 &&+\fr12 s(\z_1\cdot k_4\;\z_3\cdot k_2\; \z_2\cdot \z_4
           +\z_2\cdot k_3\;\z_4\cdot k_1\; \z_1\cdot \z_3\nn\\
   &&\quad\quad        +\z_1\cdot k_3\;\z_4\cdot k_2\; \z_2\cdot \z_3
           +\z_2\cdot k_4\;\z_3\cdot k_1\; \z_1\cdot \z_4)\nn\\
&&+\fr12 t(\z_2\cdot k_1\;\z_4\cdot k_3\; \z_3\cdot \z_1
           +\z_3\cdot k_4\;\z_1\cdot k_2\; \z_2\cdot \z_4\nn\\
   &&\quad\quad        +\z_2\cdot k_4\;\z_1\cdot k_3\; \z_3\cdot
   \z_4
           +\z_3\cdot k_1\;\z_4\cdot k_2\; \z_2\cdot \z_1)\nn\\
&& +\fr12 u(\z_1\cdot k_2\;\z_4\cdot k_3\; \z_3\cdot \z_2
           +\z_3\cdot k_4\;\z_2\cdot k_1\; \z_1\cdot \z_4\nn\\
   &&\quad\quad        +\z_1\cdot k_4\;\z_2\cdot k_3\; \z_3\cdot
   \z_4
           +\z_3\cdot k_2\;\z_4\cdot k_1\; \z_1\cdot \z_2)
           \label{k}
 \eea
 %%%
One can show that
 %%%
 \bea
 <V_gV_gV_gV_g\;V_{G,r_0^{-4}} >= \fr{4\pi g_s}{\e_y \L^4}\;K
 \eea
 %%%
 In sec 3 we illustrate the computations explicitly working out the
 coefficients of all $\z\cdot \z\;\z\cdot \z$-terms and a few
 $\z\cdot k\;\z\cdot k\;\z\cdot \z$-terms.
The next order geometry vertex operator yields vanishing expression
as in the scalar case,
%%%
 \bea
 <V_gV_gV_gV_g\;V_{G,r_0^{-5}} >=0
 \eea
 %%%
\\

The results of the two subsections above verify the conjecture at
the first two leading orders. Let's compare the results with the one
loop divergence. The one loop divergence in each case comes with a
diverging factor
 %%%
 \be
 \int_\e \fr{1}{y^2}\sim \fr1{\e}
 \ee
 %%%
This implies a relation between $\e,\e_y$ and $\L$. Up to an
immaterial numerical factor it is
 %%%
 \bea
 \fr{1}{\e_y \L^4}=\fr1{\e}
 \eea
 %%%
We now turn to the actual derivation of the results presented in
this section.

%%%%%%%%%%%%%%%%%%%%%%%%%%%%%%%%%%%%%%%%%%%%%%%%%%%%%%%%%%%%%%%%
\section{One loop divergence cancellation }
%%%%%%%%%%%%%%%%%%%%%%%%%%%%%%%%%%%%%%%%%%%%%%%%%%%%%%%%%%%%%%%%
 An M-point amplitude in
general is given by\footnote{For a review see \cite{gsw,kaku}}
 %%%
 \bea
 A_M=\int d\m
 %\int DX ds \;e^{-S}
 <\prod_{i=1}^{M}\;V(k_i)>
 \eea
 %%%
The measure $d\m$ is
 %%%
 \bea
 d\mu=|(x_1-x_2)(x_{1}-x_{M})(x_2-x_{M})|
 \int dx_3...dx_{M-1}\prod_1^{M-1}\th(x_{r
 }-x_{r+1}) \label{dmu}
 \eea
 %%%
To remove the divergence we proposed \cite{Park:2008sg} to consider
%%%
 \bea
 A_M=\int_0^1 dx\int_{x_1}^\infty dy
 %\int DX ds \;e^{-S}
 <\prod_{i=1}^{M}\;V(k_i)\;V_G(y)>
 \eea
 %%%
where $V(k_i)$ denotes an external state and $V_G$ the geometry
vertex operator. We have chosen the location of states such that
$x_4<x_3<x_2<x_1<y$ with $x_4=0,x_3=x,x_2=1$. This is a natural
choice in light of the view that $V_G|0>$ represents a some kind of
asymptotic state. At the end of each computation we take $x_1\ra
\infty$. Since the measure gives the factor $x_1^2$ one can keep
only the terms that comes with $\fr1{x_1^2}$ when computing
$<\prod_{i=1}^{M}\;V(k_i)\;V_G(y)>$. To regulate the divergence that
occurs when $y \ra x_1$ the $y$-integral range is adjusted to
 %%%
 \bea
\int_{x_1+\e_y}^\infty dy
 \eea
 %%%
We will focus on the four point amplitudes. As for the three point,
one loop, with or without the geometry vertex operator, vanishes due
to the index structures. One of the expansion parameters is taken to
be $q$,
 %%%
 \bea
 && q=\fr{Q}{r_0^4}\quad \mbox{with}\;\; Q=4\pi g^2  \a'^2
 \eea
 Note the following to keep the same
orders of the expansion parameters
 %%%
 \bea
 \pa X^+=l^2 p^+,\quad H_0=1+q,\quad H_0'=-4 \fr{q}{r_0}
 \eea
 %%%
Introducing a dimensionless constant $\L$ we measure the norm of
$r_0$ in terms of the string constant $\a'$
 %%%
 \bea
 r_0^4 =\L^4\, \a'^2
 \eea
 %%%
so that
 %%%
 \bea
 q=\fr{4\pi g^2}{\L^4}
 \eea
 %%%
The parameter $N$ that represents the number of branes will appear
through the Chan-Paton procedure.
 We take
 %%%
 \bea
 \L, g
 \eea
 %%%
as the expansion parameters for our perturbative analysis. The
bosonic and the fermionic propagators are respectively
%%%
 \bea
 <X^i X^j>&=& -2\a'\eta^{ij} \ln|x-x'| \nn\\
 <S_1^{a_1}S_1^{a_2}>&=& \fr{\d^{a_1a_2}}{x_1-x_2}
 \label{ss}
 \eea
 %%%%
 In the computations below we
 wick-rotate not only the world-sheet parameter $\t$ but also
  $\s$. The latter is implied by T-duality. The same Wick rotation
  was used in the previous work \cite{Park:2008sg}. It is useful to note that
 %%%
 \bea
 &&\Tr\; \g^{u_1v_1}\g^{u_2v_2}= -8(\d_{u_1u_2}\d_{v_1v_2}
  -\d_{u_1v_2}\d_{u_2v_1}),
  \eea
  %%%
The $\g$'s here are 8 by 8 matrices. One can easily check that
 %%%
 \bea
  <R^{u_1v_1}R^{u_2v_2}>=
  -\fr{(\d_{u_1u_2}\d_{v_1v_2}-\d_{u_1v_2}\d_{u_2v_1})}{(x_1-x_2)^2}
 \eea
 %%%
and
%%%
 \bea
&& <R^{u_1v_1}(x_1)R^{u_2v_2}(x_2)R^{u_3v_3}(x_3)> \nn\\
 =&&-\fr{1}{x_{12}x_{23}x_{13}}\;
 (\d_{u_2u_3}\d_{u_1v_2}\d_{v_1v_3}-\d_{u_2u_3}\d_{u_1v_3}\d_{v_1v_2}
  -\d_{u_2v_3}\d_{u_1v_2}\d_{v_1u_3}\nn\\
  &&\hspace{.8in} +\d_{u_2v_3}\d_{u_1u_3}\d_{v_1v_2}
  -\d_{u_3v_2}\d_{u_1u_2}\d_{v_1v_3}
  +\d_{u_3v_2}\d_{u_1v_3}\d_{u_2v_1}\nn\\
  &&\hspace{.8in}+\d_{v_2v_3}\d_{u_1u_2}\d_{v_1u_3}
  -\d_{v_2v_3}\d_{u_1u_3}\d_{u_2v_1})
 \eea
 %%%
 The product of four $R$'s can be similarly computed. The result is
 rather long so we do not present it here but refer to
 \cite{Park:2008sg}.
In many intermediate steps of the computations below momentum
conservation is used. For example in some of the correlators the
leading order term comes with $\fr{1}{x_1}$. This would lead to a
divergent result since only a factor of $x_1^2$ is present in the
integration measure. Often the term gets killed by momentum
conservation if not by its index structure. We keep $\a',l(\equiv
\sqrt{2\a'})$ in some places but in others we have used their
explicit values,
 %%%
 \bea
 \a'=\fr12,\quad l=1
 \eea
 %%%

%%%%%%%%%%%%%%%%%%%%%%%%%%%%%%%%%%%%%%%%%%%%%%%%%%%%%%%%%%%%%%%%
\subsection{scalar scattering case}
%%%%%%%%%%%%%%%%%%%%%%%%%%%%%%%%%%%%%%%%%%%%%%%%%%%%%%%%%%%%%%%%

For convenience we record the explicit form of the product of four
scalar vertex operators
 %%%
 \bea
&& V^{m_1}_s(x_1) V^{m_2}_s(x_2) V^{m_3}_s(x_3) V^{m_4}_s(x_4) \nn\\
&& =X'^{m_1}X'^{m_2}X'^{m_3}X'^{m_4}
+l^8R^{m_1v_1}k_1^{v_1}R^{m_2v_2}k_2^{v_2}R^{m_3v_3}k_3^{v_3}R^{m_4v_4}k_4^{v_4}\nn\\
 %%%%%%%%
&&-l^2\left[X'^{m_1}X'^{m_2}X'^{m_3}R^{m_4v_4}k_4^{v_4}
  +X'^{m_1}X'^{m_2}X'^{m_4}R^{m_3v_3}k_4^{v_3}\right.\nn\\
&&\left.\;\;\;\;+X'^{m_1}X'^{m_3}X'^{m_4}R^{m_2v_2}k_2^{v_2}
   +X'^{m_2}X'^{m_3}X'^{m_4}R^{m_1v_1}k_1^{v_1}\right]\nn\\
 %%%%%%%%%
&&+l^4\left[X'^{m_1}X'^{m_2}R^{m_3v_3}k_3^{v_3}R^{m_4v_4}k_4^{v_4}
  +X'^{m_3}X'^{m_4}R^{m_1v_1}k_1^{v_1}R^{m_2v_2}k_2^{v_2}\right.\nn\\
 && \left.\;\;\;\;+X'^{m_1}X'^{m_4}R^{m_2v_2}k_2^{v_2}R^{m_3v_3}k_3^{v_3}
  +X'^{m_1}X'^{m_3}R^{m_2v_2}k_2^{v_2}R^{m_4v_4}k_4^{v_4}\right.\nn\\
  &&\left. \;\;\;\;+X'^{m_2}X'^{m_3}R^{m_1v_1}k_1^{v_1}R^{m_4v_4}k_4^{v_4}
  +X'^{m_2}X'^{m_4}R^{m_1v_1}k_1^{v_1}R^{m_3v_3}k_3^{v_3}\right]
  \nn\\
 %%%%%%%%
 &&-l^6\left[X'^{m_1}R^{m_2v_2}k_2^{v_2}R^{m_3v_3}k_3^{v_3}R^{m_4v_4}k_4^{v_4}
        +X'^{m_2}R^{m_1v_1}k_1^{v_1}R^{m_3v_3}k_3^{v_3}R^{m_4v_4}k_4^{v_4}
        \right. \nn\\
&&\left.\;\;\;\;+X'^{m_3}R^{m_1v_1}k_1^{v_1}R^{m_2v_2}k_2^{v_2}R^{m_4v_4}k_4^{v_4}
        +X'^{m_4}R^{m_1v_1}k_1^{v_1}R^{m_2v_2}k_2^{v_2}R^{m_3v_3}k_3^{v_3}
        \right]
 \eea
 %%%
The form is appropriate before the Wick rotation which will be taken
into account in each individual computation below. With it we
multiply the geometry vertex operator at each order, such as
$V_{G,r_0^{-4}},V_{G,r_0^{-5}}\;\; \mbox{etc}$, and compute the
resulting correlator. Each level geometry vertex operator has
several terms: we compute them one by one and put the results
together at the end. A pattern emerges on how the desired kinematic
structure arises: only a first few leading terms are responsible for
the structure with the higher order terms yielding vanishing
results.

%%%%%%%%%%%%%%%%%%%%%%%%%%%%%%%%%%%%%
\subsubsection{leading order computation }
%%%%%%%%%%%%%%%%%%%%%%%%%%%%%%%%%%%%%
The leading vertex operator is given by
%%%
 \bea
\pi V_{G,r_0^{-4}}
 && \simeq  \;\fr{q}4\sqrt{h}\;h^{ij}
 \left(\pa_i X^u \pa_j X^v \eta_{uv} -
 \pa_i X^m \pa_j X^n \eta_{mn} \right)
   +\fr{il^2\,q}{4}
   (\sqrt{h}\;h^{0j}+\ve^{0j})
         (S\pa_jS)\nn\\
  %%%%%%%%%%%%%%%%%%%%%%%%%%%%%%%%%%
 %%%%%%%%%%%%%%%%%%%%%%%%%%%%%%%%
 && \simeq \fr{q}4  \;
 \left(-\pa_\s X^m \pa_\s X^m-\pa_\t X^u \pa_\t X^u
  +il^2\,(- S\pa_\t S- S\pa_\s S )\right) \label{leadingV3}
  \eea
  %%%
When one goes from the first to second one drops certain bosonic
terms because of $\s=0$. With the first correlator
 \[<V^{m_1}_s(x_1) V^{m_2}_s(x_2) V^{m_3}_s(x_3)
V^{m_4}_s(x_4)\;\pa_i X^m \pa_j X^n \eta_{mn}>\], certain terms drop
either because of the dimensional regularization or/and they contain
an odd number of $X^m$ fields:
 %%%
 \bea
&&<V^{m_1}_s(x_1) V^{m_2}_s(x_2) V^{m_3}_s(x_3)
V^{m_4}_s(x_4)\;\pa_i X^m(y) \pa_j X^n(y) \eta_{mn}> \nn\\
&& =<X'^{m_1}X'^{m_2}X'^{m_3}X'^{m_4}\;\pa_i X^m \pa_j X^n
\eta_{mn}>
\nn\\
 %%%%%%%%%
&&-l^4<\left[X'^{m_1}X'^{m_2}R^{m_3v_3}k_3^{v_3}R^{m_4v_4}k_4^{v_4}
  +X'^{m_3}X'^{m_4}R^{m_1v_1}k_1^{v_1}R^{m_2v_2}k_2^{v_2}\right.\nn\\
 && \left.\;\;\;\;+X'^{m_1}X'^{m_4}R^{m_2v_2}k_2^{v_2}R^{m_3v_3}k_3^{v_3}
  +X'^{m_1}X'^{m_3}R^{m_2v_2}k_2^{v_2}R^{m_4v_4}k_4^{v_4}\right.\\
  &&\left. \;\;\;\;+X'^{m_2}X'^{m_3}R^{m_1v_1}k_1^{v_1}R^{m_4v_4}k_4^{v_4}
  +X'^{m_2}X'^{m_4}R^{m_1v_1}k_1^{v_1}R^{m_3v_3}k_3^{v_3}\right]
  \;\pa_i X^m \pa_j X^n \eta_{mn}>\nn
 \eea
 %%%
 The signs of $l^4$-terms have been flipped by Wick rotation
 in the $\s$ direction, which is implied by T-duality. Combining the two contributions
  one gets after collecting terms of the same types
 %%%
 \bea
 &&\int_0^1 dx \int_{x_1+\e_y}^\infty dy\;<V^{m_1}_s(x_1) V^{m_2}_s(x_2) V^{m_3}_s(x_3)
V^{m_4}_s(x_4)\;\pa_i X^m(y) \pa_j X^n(y) \eta_{mn}> \nn\\
%%%%%%%%%%%%%%%%%%%%%%%%%%%%%%%%%%%%%%%
=  &&\int_0^1 dx\left[-\fr{16}{{\epsilon }_4}\left( {{\xi }_1}\cdot
{{\xi }_4}\,{{\xi }_2}\cdot {{\xi }_3}\,
     \frac{1}{\left( -1 + x \right)^2}
      +{{{\xi }_1}\cdot {{\xi }_3}\,{{\xi }_2}\cdot {{\xi}_4}}
 + \frac{{{\xi }_1}\cdot {{\xi }_2}
 \,{{\xi }_3}\cdot {{\xi }_4}}{x^2\,} \right) \,{\alpha '}^3\right.
 \nn\\
&&\left. -\fr{16}{{\epsilon }_4}\left( \frac{t\,{{\xi }_1}\cdot
{{\xi }_4}\,{{\xi }_2}\cdot {{\xi }_3}}
     {{\left( -1 + x \right) }^2\,}
     +{u\,{{\xi }_1}\cdot {{\xi }_3}\,{{\xi }_2}\cdot {{\xi }_4}}
    +\frac{s\,{{\xi }_1}\cdot {{\xi }_2}\,{{\xi }_3}\cdot {{\xi
}_4}}{x^2\,} \right) \,{\alpha '}^4\;\right]
 \eea
 %%%
After performing the x-integration it reproduces the tree result of
the four point amplitude without the geometry vertex operator,
%%%
 \bea
\fr1{\e_y}\; \fr1{4 }(su\;\xi_1\cdot \xi_4\;\xi_2\cdot \xi_3
 +tu\;\xi_1\cdot \xi_2\;\xi_3\cdot \xi_4
 +st\;\xi_2\cdot \xi_4\;\xi_1\cdot \xi_3)
 \eea
 %%%
 The next correlator to consider is
\[<V^{m_1}_s(x_1) V^{m_2}_s(x_2) V^{m_3}_s(x_3)
V^{m_4}_s(x_4)\;\pa_i X^u \pa_j X^v \eta_{uv}>\]
 Here again certain terms drop trivially
either because of the dimensional regularization or/and because they
contain an odd number of $X^m$ fields:
 %%%
 \bea
&&<V^{m_1}_s(x_1) V^{m_2}_s(x_2) V^{m_3}_s(x_3)
V^{m_4}_s(x_4)\;\pa_i X^u(y) \pa_j X^v(y) \eta_{uv}> \nn\\
&& =<X'^{m_1}X'^{m_2}X'^{m_3}X'^{m_4}\;\pa_i X^u \pa_j X^v
\eta_{uv}>
\nn\\
%%%%%%
 &&+l^8<R^{m_1v_1}k_1^{v_1}R^{m_2v_2}k_2^{v_2}R^{m_3v_3}k_3^{v_3}R^{m_4v_4}k_4^{v_4}
  \;\pa_i X^u\pa_j X^v \eta_{uv}>\nn\\
 %%%%%%%%%
&&-l^4<\left[X'^{m_1}X'^{m_2}R^{m_3v_3}k_3^{v_3}R^{m_4v_4}k_4^{v_4}
  +X'^{m_3}X'^{m_4}R^{m_1v_1}k_1^{v_1}R^{m_2v_2}k_2^{v_2}\right.\nn\\
 && \left.\;\;\;\;+X'^{m_1}X'^{m_4}R^{m_2v_2}k_2^{v_2}R^{m_3v_3}k_3^{v_3}
  +X'^{m_1}X'^{m_3}R^{m_2v_2}k_2^{v_2}R^{m_4v_4}k_4^{v_4}\right.\\
  &&\left. \;\;\;\;+X'^{m_2}X'^{m_3}R^{m_1v_1}k_1^{v_1}R^{m_4v_4}k_4^{v_4}
  +X'^{m_2}X'^{m_4}R^{m_1v_1}k_1^{v_1}R^{m_3v_3}k_3^{v_3}\right]
  \;\pa_i X^u \pa_j X^v \eta_{uv}>\nn
 \eea
 %%%
 %{\bf where the signs of the $l^4$-terms have been flipped.}
After some algebra one can show that
 %%%
 \bea
&&<V^{m_1}_s(x_1) V^{m_2}_s(x_2) V^{m_3}_s(x_3)
V^{m_4}_s(x_4)\;\pa_i X^u(y) \pa_j X^v(y) \eta_{uv}>\;\;\sim
\fr{1}{x_1^3}
 \eea
 %%%
It vanishes as $x_1\ra \infty$ since it still goes $\sim\fr1{x_1}$
even after taking the measure into account. As a matter of fact many
terms in the higher-order $V_G$ seem to vanish for the same reason.
The last correlator which is with $S(y)\pa S(y)$ vanishes due to the
fermionic equation of motion. This completes the proof that at the
leading order in $r_0$-expansion the geometry vertex operator does
produce the correct structure to cancel the one-loop divergence.

%%%%%%%%%%%%%%%%%%%%%%%%%%%%%%%%%%%%%
\subsubsection{next leading order computation}
%%%%%%%%%%%%%%%%%%%%%%%%%%%%%%%%%%%%%
 In the next leading order the geometry vertex operator
 is\footnote{There are terms that are of the form $\pa X\pa X X$ that come from
 the first line of (\ref{startingactionfinal}) after expanding $H^{\pm 1/2}$. They
 trivially vanish due to the dimensional regularization and/or their index
 structures.  }
 %%%
 \bea
\pi V_{G,r_0^{-5}}=&& -\fr{i}4 (\sqrt{h}h^{ij}-\ve^{ij})\pa_i X^+
H_0^{-5/4}\fr{H_0'}{r_0}
       \pa_jX^m X_0^n\;(S\g^{mn}S)\nn\\
 &&+\fr{i}4 (\sqrt{h}h^{ij}-\ve^{ij})\pa_i X^+ H_0^{-7/4}\fr{H_0'}{r_0} \pa_jX^u
        X_0^n\; (S\g^{un}S)
 \eea
 %%%
As before several terms drop trivially
 %%%
 \bea
&&<V^{m_1}_s(x_1) V^{m_2}_s(x_2) V^{m_3}_s(x_3)
V^{m_4}_s(x_4)\;\pa_jX^m (S\g^{mn}S >X_0^n \nn\\
=&&<( -l^2\left[X'^{m_1}X'^{m_2}X'^{m_3}R^{m_4v_4}k_4^{v_4}
  +X'^{m_1}X'^{m_2}X'^{m_4}R^{m_3v_3}k_4^{v_3}\right.\nn\\
&&\left.\;\;\;\;+X'^{m_1}X'^{m_3}X'^{m_4}R^{m_2v_2}k_2^{v_2}
   +X'^{m_2}X'^{m_3}X'^{m_4}R^{m_1v_1}k_1^{v_1}\right]\nn\\
 %%%%%%%%
 &&-l^6\left[X'^{m_1}R^{m_2v_2}k_2^{v_2}R^{m_3v_3}k_3^{v_3}R^{m_4v_4}k_4^{v_4}
        +X'^{m_2}R^{m_1v_1}k_1^{v_1}R^{m_3v_3}k_3^{v_3}R^{m_4v_4}k_4^{v_4}
        \right. \nn\\
&&\left.\;\;\;\;+X'^{m_3}R^{m_1v_1}k_1^{v_1}R^{m_2v_2}k_2^{v_2}R^{m_4v_4}k_4^{v_4}
        +X'^{m_4}R^{m_1v_1}k_1^{v_1}R^{m_2v_2}k_2^{v_2}R^{m_3v_3}k_3^{v_3}
        \right])\nn\\
      && \pa_jX^m (S\g^{mn}S >X_0^n\nn\\
 %%%%%%%%%%%%%%%%%%%%%%
 =&&<(-l^6\left[X'^{m_1}R^{m_2v_2}k_2^{v_2}R^{m_3v_3}k_3^{v_3}R^{m_4v_4}k_4^{v_4}
        +X'^{m_2}R^{m_1v_1}k_1^{v_1}R^{m_3v_3}k_3^{v_3}R^{m_4v_4}k_4^{v_4}
        \right. \nn\\
&&\left.\;\;\;\;+X'^{m_3}R^{m_1v_1}k_1^{v_1}R^{m_2v_2}k_2^{v_2}R^{m_4v_4}k_4^{v_4}
        +X'^{m_4}R^{m_1v_1}k_1^{v_1}R^{m_2v_2}k_2^{v_2}R^{m_3v_3}k_3^{v_3}
        \right])\nn\\
      && \pa_jX^m (S\g^{mn}S >X_0^n
 \eea
 %%%
One can see that the $l^2$-terms vanish due to the index structure
as follows. For example consider $<R^{m_1v_1}\,S\g^{mn}S>\sim
(\d_{m_1m}\d_{v_1n}-\d_{m_1n}\d_{v_1m})=0$. (Recall that the $m$ or
$n$ indices run in the transverse directions whereas the $u$ or $v$
in the longitudinal space.) The $l^6$-terms also vanish for the same
reason. Therefore at this order one gets
%%%
 \bea
 <V_sV_sV_sV_s\; V_{G,r_0^{-5}}) > =0
 \eea
 %%%

%%%%%%%%%%%%%%%%%%%%%%%%%%%%%%%%%%%%%%%%%%%%%%%%%%%%%%%%%%%%%%%%
\subsection{vector scattering case}
%%%%%%%%%%%%%%%%%%%%%%%%%%%%%%%%%%%%%%%%%%%%%%%%%%%%%%%%%%%%%%%%
The explicit form of the product of four vector vertex operators is
 %%%
 \bea
&& V^{u_1}_g(x_1) V^{u_2}_g(x_2) V^{u_3}_g(x_3) V^{u_4}_g(x_4) \nn\\
&& =\dot{X}^{u_1}\dot{X}^{u_2}\dot{X}^{u_3}\dot{X}^{u_4}
+l^8R^{u_1v_1}k_1^{v_1}R^{u_2v_2}k_2^{v_2}R^{u_3v_3}k_3^{v_3}R^{u_4v_4}k_4^{v_4}\nn\\
 %%%%%%%%
&&-l^2\left[\dot{X}^{u_1}\dot{X}^{u_2}\dot{X}^{u_3}R^{u_4v_4}k_4^{v_4}
  +\dot{X}^{u_1}\dot{X}^{u_2}\dot{X}^{u_4}R^{u_3v_3}k_4^{v_3}\right.\nn\\
&&\left.\;\;\;\;+\dot{X}^{u_1}\dot{X}^{u_3}\dot{X}^{u_4}R^{u_2v_2}k_2^{v_2}
   +\dot{X}^{u_2}\dot{X}^{u_3}\dot{X}^{u_4}R^{u_1v_1}k_1^{v_1}\right]\nn\\
 %%%%%%%%%
&&+l^4\left[\dot{X}^{u_1}\dot{X}^{u_2}R^{u_3v_3}k_3^{v_3}R^{u_4v_4}k_4^{v_4}
  +\dot{X}^{u_3}\dot{X}^{u_4}R^{u_1v_1}k_1^{v_1}R^{u_2v_2}k_2^{v_2}\right.\nn\\
 && \left.\;\;\;\;+\dot{X}^{u_1}\dot{X}^{u_4}R^{u_2v_2}k_2^{v_2}R^{u_3v_3}k_3^{v_3}
  +\dot{X}^{u_1}\dot{X}^{u_3}R^{u_2v_2}k_2^{v_2}R^{u_4v_4}k_4^{v_4}\right.\nn\\
  &&\left. \;\;\;\;+\dot{X}^{u_2}\dot{X}^{u_3}R^{u_1v_1}k_1^{v_1}R^{u_4v_4}k_4^{v_4}
  +\dot{X}^{u_2}\dot{X}^{u_4}R^{u_1v_1}k_1^{v_1}R^{u_3v_3}k_3^{v_3}\right]
  \nn\\
 %%%%%%%%
 &&-l^6\left[\dot{X}^{u_1}R^{u_2v_2}k_2^{v_2}R^{u_3v_3}k_3^{v_3}R^{u_4v_4}k_4^{v_4}
        +\dot{X}^{u_2}R^{u_1v_1}k_1^{v_1}R^{u_3v_3}k_3^{v_3}R^{u_4v_4}k_4^{v_4}
        \right. \nn\\
&&\left.\;\;\;\;+\dot{X}^{u_3}R^{u_1v_1}k_1^{v_1}R^{u_2v_2}k_2^{v_2}R^{u_4v_4}k_4^{v_4}
        +\dot{X}^{u_4}R^{u_1v_1}k_1^{v_1}R^{u_2v_2}k_2^{v_2}R^{u_3v_3}k_3^{v_3}
        \right] \label{startingaction-vector}
 \eea
 %%%
which is before the Wick rotation.

%%%%%%%%%%%%%%%%%%%%%%%%%%
 \subsubsection{leading order computation}
%%%%%%%%%%%%%%%%%%%%%%%%
Since the large $r_0$-expansion is in terms of the transverse
coordinates $X^m$ and the transverse and the longitudinal
coordinates do not contract each other one does not have to perform
the expansion when considering purely vector state
scattering.\footnote{This is true for the pure gauge boson
scattering. Once one puts the gaugino state it will not be so in
general since it contains $X'$ as well as ${\dot{X}}$.} To cancel
the divergence of the four vector scattering the relevant terms of
the vertex operator $V_G$ are
%%%
 \bea
\pi V_G
 \Rightarrow && \;-\!\!\fr{1}2\sqrt{h}\;h^{ij}
 \left(\pa_i X^u \pa_j X^v \eta_{uv} (H_0^{-1/2}-1)
 %+\pa_i X^m \pa_j X^n \eta_{mn} (H_0^{1/2}-1)
 \right)
  \nn\\
 %%%%%%%%%%%%%%%%%%%%%%%%%%%%%%%%
 && +\fr1{2p^+} \{ \nn\\
  && -2i(\sqrt{h}\;h^{ij}-\ve^{ij})\pa_i X^+ (H_0^{-1/4}-1)
         (S\pa_jS)  \nn\\
 &&+\fr{i}4 (\sqrt{h}h^{ij}-\ve^{ij})\pa_i X^+ H_0^{-7/4}\fr{H_0'}{r_0} \pa_jX^u
        X_0^m\; (S\g^{um}S)\nn\\
  % &&    -\fr{i}4 (\sqrt{h}h^{ij}+\ve^{ij})\pa_i X^+ H^{-5/4}\fr{H'}{r}
  %     \pa_jX^m X^n\;(S\g^{mn}S)\nn\\
        && \quad\quad\hspace{1in}\} \nn\\
 %%%%%%%%%%%%%%%%%%%%%%%%%%%%%%%%%%
       &&+\fr1{4(p^+)^2}\sqrt{h}h^{ij}\pa_i X^+\pa_j X^+\;H_0^{-1/2}\{\nn\\
  &&-\fr{17}{1536}\k_{10}(S\g^{uv} S)( S\g^{uv} S)+\left[
  \fr{43}{768}\k_{10}+\fr1{192}\k_{20}
  \right](S\g^{au} S)( S\g^{au} S) \nn\\
  &&-\left[
  \fr1{192}\k_{20} +\fr1{128}\k_{10}\right] (S\g^{ab} S)( S\g^{ab} S)\nn\\
  &&+X_0^aX_0^b
  \fr{1}{r_0^2}\left[\fr{31}{768}\k_{10}-\fr1{32}\k_{20}
  \right](S\g^{au} S)( S\g^{bu} S)\nn\\
  &&+X_0^aX_0^b \fr{1}{r_0^2}\left[
  +\fr1{32}\k_{20}+\fr{29}{384}\k_{10}\right](S\g^{ac} S)( S\g^{bc} S) \nn\\
 &&\}
 \eea
 %%%
 A few of the $S$-quadratic terms have been dropped because it does not make
 any contribution in the dimensional regularization. The leading
 order operator is
 %%%
 \bea
\pi V_{G,r_0^{-4}}
 && \Rightarrow \fr{q}4  \;
 \left(-\pa_\t X^u \pa_\t X^u
  +il^2\,(- S\pa_\t S- S\pa_\s S )\right)
  \eea
  %%%
 As for the following correlators with $(- S\pa_\t S- S\pa_\s S )$
it vanishes because of the fermionic field equation. Some of the
terms are trivially zero. The $<RRRR\pa X\pa X>$ goes as $1/x_1^4$
so it lead to a vanishing result, so one can compute
 %%%
 \bea
 &&<(\dot{X}^{u_1}\dot{X}^{u_2}\dot{X}^{u_3}\dot{X}^{u_4}
\nn\\
 %%%%%%%%%
&&+l^4\left[\dot{X}^{u_1}\dot{X}^{u_2}R^{u_3v_3}k_3^{v_3}R^{u_4v_4}k_4^{v_4}
  +\dot{X}^{u_3}\dot{X}^{u_4}R^{u_1v_1}k_1^{v_1}R^{u_2v_2}k_2^{v_2}\right.\nn\\
 && \left.\;\;\;\;+\dot{X}^{u_1}\dot{X}^{u_4}R^{u_2v_2}k_2^{v_2}R^{u_3v_3}k_3^{v_3}
  +\dot{X}^{u_1}\dot{X}^{u_3}R^{u_2v_2}k_2^{v_2}R^{u_4v_4}k_4^{v_4}\right.\nn\\
  &&\left. \;\;\;\;+\dot{X}^{u_2}\dot{X}^{u_3}R^{u_1v_1}k_1^{v_1}R^{u_4v_4}k_4^{v_4}
  +\dot{X}^{u_2}\dot{X}^{u_4}R^{u_1v_1}k_1^{v_1}R^{u_3v_3}k_3^{v_3}\right]
  \nn\\
 %%%%%%%%
 &&-l^6\left[\dot{X}^{u_1}R^{u_2v_2}k_2^{v_2}R^{u_3v_3}k_3^{v_3}R^{u_4v_4}k_4^{v_4}
        +\dot{X}^{u_2}R^{u_1v_1}k_1^{v_1}R^{u_3v_3}k_3^{v_3}R^{u_4v_4}k_4^{v_4}
        \right. \nn\\
&&\left.\;\;\;\;+\dot{X}^{u_3}R^{u_1v_1}k_1^{v_1}R^{u_2v_2}k_2^{v_2}R^{u_4v_4}k_4^{v_4}
        +\dot{X}^{u_4}R^{u_1v_1}k_1^{v_1}R^{u_2v_2}k_2^{v_2}R^{u_3v_3}k_3^{v_3}
        \right])\nn\\
&&      \pa_i X^u \pa_j X^v \eta_{uv}  >
 \eea
 %%%
The task is to reproduce the kinematic structure given in (\ref{k}).
We take a few examples to illustrate how things work. As for the
coefficient of $\z_1\cdot \z_2\;\z_3\cdot \z_4$ only $XXXX\pa X\pa
X$ and $XXRR\pa X\pa X$ contribute and one gets
 %%%
 \bea
 \frac{{{\z }_1}\cdot {{\z }_2}
 \,{{\z }_3}\cdot {{\z }_4}}{x^2\,} \left( -\fr{16}{{\epsilon }_4}
  \,{\alpha '}^3
 -\fr{4}{{\epsilon }_4}
    \,{\alpha '}^2\right)
 \eea
 %%%
 Note that we sometimes use $\a'=1/2$ here and there so the powers
of $\a'$ is not systematic. After the x-integration one gets the
expected result. Let's consider an example of the form $\z\cdot
k\;\z\cdot k\;\z\cdot \z$ The coefficient of $\z_1\cdot \z_2$ comes
from $XXXX\pa X\pa X$ and $XXRR\pa X\pa X$:
 %%%
 \bea
 && 32\left( \frac{{k_2}\cdot {{\zeta }_3}\,{k_2}\cdot {{\zeta }_4}}{-1 + x}
 +\frac{{k_2}\cdot {{\zeta }_3}\,{k_3}\cdot {{\zeta }_4}}{\left(
  -1 + x \right) \,x}
  +\frac{{k_2}\cdot {{\zeta }_4}\,{k_4}\cdot {{\zeta }_3}}{x}
 +\frac{{k_3}\cdot {{\zeta }_4}\,{k_4}\cdot {{\zeta }_3}}{x^2}
\right) \,\,
    \frac{{\alpha '}^4}{{\epsilon }_4}\nn\\
&&-\frac{8\,{{\zeta }_3}\cdot {k_4}\,{{\zeta }_4}\cdot
{k_3}\,{\alpha '}^2}{x^2\,{{\epsilon }_4}}
 \eea
 %%%
which leads, after the x-integration, to the correct result of
 %%%
 \bea
 u \;\z_3\cdot k_2\; \z_4 \cdot k_1+ t \;\z_3\cdot k_1\; \z_4 \cdot k_2
 \eea
 %%%
 Even for the same $\z\cdot
k\;\z\cdot k\;\z\cdot \z$-type terms the conspiracy between the
intermediate terms can be different as can be seen in the
computation of $\z_2\cdot \z_4$-term. Here all three different type
of terms contribute.
 %%%
 \bea
 &&<XXXX\pa X \pa X>\nn\\
&\Rightarrow& \frac{32}{{{\epsilon }_4}}\left( \frac{-\,{{\zeta
}_1}\cdot {k_2}\,{{\zeta }_3}\cdot {k_2}}{-1 + x} -
      \frac{\,x\,{{\zeta }_1}\cdot {k_3}\,{{\zeta }_3}\cdot {k_2}}{-1 + x} -
      \frac{\,{{\zeta }_1}\cdot {k_2}\,{{\zeta }_3}\cdot {k_4}}{-1 + x}
  \right. \nn\\
   &&\left. \hspace{1in}  +\frac{\,{{\zeta }_1}\cdot {k_2}\,{{\zeta }_3}\cdot {k_4}}
      {\left( -1 + x \right) \,x} -
      \,{{\zeta }_1}\cdot {k_3}\,{{\zeta }_3}\cdot {k_4} \right)
      \,{\alpha '}^4
 \nn\\
 &&-<XXRR\pa X \pa X>\nn\\
&\Rightarrow &
 -\frac{{\alpha '}^3}{{{\epsilon }_4}}
  \left( \frac{8\,u\,{{\zeta }_1}\cdot {k_2}\,{{\zeta }_3}\cdot {k_2}}{-1 + x} +
      \frac{8\,u\,x\,{{\zeta }_1}\cdot {k_3}\,{{\zeta }_3}\cdot {k_2}}{-1 + x} +
      \frac{8\,u\,{{\zeta }_1}\cdot {k_2}\,{{\zeta }_3}\cdot {k_4}}{-1 + x}
        \right.\nn\\
     &&\left. -\frac{8\,u\,{{\zeta }_1}\cdot {k_2}\,{{\zeta }_3}\cdot {k_4}}
      {\left( -1 + x \right) \,x}
     -\frac{8\,u\,{{\zeta }_1}\cdot {k_3}\,{{\zeta }_3}\cdot {k_4}}{-1 +x}
     + \frac{8\,u\,x\,{{\zeta }_1}\cdot {k_3}\,{{\zeta }_3}\cdot {k_4}}{-1 + x} \right)
 \nn\\
 && <XXRR\pa X \pa X>\nn\\
 &\Rightarrow &
 \frac1{{{\epsilon }_4}}\left( \frac{-i \,s\,{{\zeta }_1}\cdot {k_2}
 \,{{\zeta }_3}\cdot {k_2}}
       {\left( -1 + x \right) \,x}
       - \frac{i \,s\,{{\zeta }_1}\cdot {k_3}\,{{\zeta }_3}\cdot {k_2}}{-1 + x} +
      \frac{i \,t\,{{\zeta }_1}\cdot {k_2}\,{{\zeta }_3}\cdot {k_4}}
      {\left( -1 + x \right) \,x} \right.\nn\\
   &&\left.  \hspace{1in}   +
      \frac{i \,t\,{{\zeta }_1}\cdot {k_3}\,{{\zeta }_3}\cdot {k_4}}{-1 + x}
       \right)
 \eea
 %%%
One can easily show by doing the x-integration that
 %%%
 \bea
 &&<XXXX\pa X \pa X>-<XXRR\pa X \pa X>-i<XRRR\pa X \pa X>\nn\\
 \Rightarrow && s\;\z_1\cdot k_4\;\z_3\cdot k_2+t\;\z_1\cdot k_2\; \z_3\cdot k_4
 \eea
 %%%
where the Wick rot has been taken into account. The result is as
expected. Each correlator above produces many unwanted terms of
different structure, i.e., more ``mixed" types of terms. They are
combined to cancel among themselves. The details go as follows.
%%%
 \bea
 &&<XXXX\pa X \pa X>-<XXRR\pa X \pa X>-i<XXRR\pa X \pa X>\nn\\
 \Rightarrow &&
 %%%%%%%%%%%%%%%%%%%%%%%%5
\fr1{\epsilon_4}\left[\frac{2}{{\left( -1 + x \right) }^2\,x^2\,}
      \left( {{\zeta }_1}\cdot {k_2}
    + x\,{{\zeta }_1}\cdot {k_3}\right) \,
    \left( -{{\zeta }_2}\cdot {k_3} - {{\zeta }_2}\cdot {k_4} + x\,
    {{\zeta }_2}\cdot {k_4} \right) \right.\nn\\
 &&\left. \hspace{1in}  \left( x\,{{\zeta }_3}\cdot {k_2}
 - {{\zeta }_3}\cdot {k_4} + x\,{{\zeta }_3}\cdot {k_4} \right) \,
    \left( x\,{{\zeta }_4}\cdot {k_2} + {{\zeta }_4}\cdot {k_3}
    \right)\fr{}{}\right]\nn\\
    %%%%%%%%%%%%%%%%%%%%%%%%%%%%%%%%
 &&  - \fr1{\epsilon_4}\left[\frac{2}{{\left( -1 + x \right) }^2\,x^2\,
    }  \left( {{\zeta }_1}\cdot {k_2}
    + x\,{{\zeta }_1}\cdot {k_3}\right) \,
    \left( -{{\zeta }_2}\cdot {k_3} - {{\zeta }_2}\cdot {k_4} + x\,
    {{\zeta }_2}\cdot {k_4} \right) \right.\nn\\
 &&\left. \hspace{1in}  \left( x\,{{\zeta }_3}\cdot {k_2}
 - {{\zeta }_3}\cdot {k_4} + x\,{{\zeta }_3}\cdot {k_4} \right) \,
    \left( x\,{{\zeta }_4}\cdot {k_2} + {{\zeta }_4}\cdot {k_3}
    \right)\right.\nn\\
    &&+\left. \frac{\left( 2  \right) \,\left( {{\zeta }_1}\cdot
 {k_2} + x\,{{\zeta }_1}\cdot {k_3} \right) \,
    \left( {{\zeta }_2}\cdot {k_3}\,{{\zeta }_3}\cdot {k_4}\,
    {{\zeta }_4}\cdot {k_2} -
      {{\zeta }_2}\cdot {k_4}\,{{\zeta }_3}\cdot {k_2}\,
      {{\zeta }_4}\cdot {k_3} \right) }{
    \left( -1 + x \right) \,x}
    \right]\nn\\
  %%%%%%%%%%%%%%%%%%%%%%%%%%%%%%%%%%%%
  &&-\fr{i}{\epsilon_4}\left[
  \frac{\left( 2i  \right)\,\left( {{\zeta }_1}\cdot {k_2}
  + x\,{{\zeta }_1}\cdot {k_3} \right) \,
    \left( {{\zeta }_2}\cdot {k_3}\,{{\zeta }_3}\cdot {k_4}\,
    {{\zeta }_4}\cdot {k_2} -
      {{\zeta }_2}\cdot {k_4}\,{{\zeta }_3}\cdot {k_2}\,
      {{\zeta }_4}\cdot {k_3} \right) }
      {\left( -1 + x \right) \,x}
  \right]\nn\\
  &&=0
 \eea
 %%%

%%%%%%%%%%%%%%%%%%%%%%%%%%
 \subsubsection{next leading order computation}
%%%%%%%%%%%%%%%%%%%%%%%%
For the vector scattering first consider $<V_gV_gV_gV_g\; \pa_jX^m
(S\g^{mn}S) > X_0^n$. By careful inspection of the indices one can
show that it vanishes,
%%%
 \bea
 <V_gV_gV_gV_g\; \pa_jX^m (S\g^{mn}S) > X_0^n=0
 \eea
 %%%
The second term in the geometry vertex operator gives
 %%%
 \bea
 &&<(
l^8R^{u_1v_1}k_1^{v_1}R^{u_2v_2}k_2^{v_2}R^{u_3v_3}k_3^{v_3}R^{u_4v_4}k_4^{v_4}\nn\\
 %%%%%%%%%
&&+l^4\left[\dot{X}^{u_1}\dot{X}^{u_2}R^{u_3v_3}k_3^{v_3}R^{u_4v_4}k_4^{v_4}
  +\dot{X}^{u_3}\dot{X}^{u_4}R^{u_1v_1}k_1^{v_1}R^{u_2v_2}k_2^{v_2}\right.\nn\\
 && \left.\;\;\;\;+\dot{X}^{u_1}\dot{X}^{u_4}R^{u_2v_2}k_2^{v_2}R^{u_3v_3}k_3^{v_3}
  +\dot{X}^{u_1}\dot{X}^{u_3}R^{u_2v_2}k_2^{v_2}R^{u_4v_4}k_4^{v_4}\right.\nn\\
  &&\left. \;\;\;\;+\dot{X}^{u_2}\dot{X}^{u_3}R^{u_1v_1}k_1^{v_1}R^{u_4v_4}k_4^{v_4}
  +\dot{X}^{u_2}\dot{X}^{u_4}R^{u_1v_1}k_1^{v_1}R^{u_3v_3}k_3^{v_3}\right]
  )\nn\\
&&      \pa_jX^u \; (S\g^{un}S)  > X_0^n
 \eea
 %%%
By inspecting the index structures again it is not difficult to tell
that the above correlators vanish: basically because all the indices
are $(u,v)$ except one which is $n$. Therefore at this order
%%%
 \bea
 <V_gV_gV_gV_g\; V_{G,r_0^{-5}}) > =0
 \eea
 %%%

%%%%%%%%%%%%%%%%%%%%%%%%%%%%%%%%%%%%%%%%%%%%%%%%%%%%%%%%%%%%%%%%
\section{Discussion and future directions}
%%%%%%%%%%%%%%%%%%%%%%%%%%%%%%%%%%%%%%%%%%%%%%%%%%%%%%%%%%%%%%%%

In this work we have shown\footnote{For a D0 brane or D1 brane it is
necessary to consider the recoil effect that was discussed for
example in \cite{Craps:2006vx}.} at the first two leading orders in
the large-$r_0$ expansion that the counter vertex operator
(\ref{startingactionfinal}) does produce the required structure
(without any extra unwanted terms) to absorb the one-loop
divergence. It is, therefore, verification of the conjecture put
forward in \cite{Park:2008sg} at the specified orders. It is
encouraging that it is possible to absorb the divergence within the
pure open string frame-work. For one thing it is not very clear how
to produce the open string kinematic factor using some kind of
explicit closed string degrees of freedom. Also even within the open
string frame-work it is a priori never guaranteed that the
computation will yield the right and only the right types of terms.

As the order increases, more\footnote{However, only the finite
number of terms contribute as mentioned previously. } terms in the
geometry vertex operator become relevant. With each term the number
of the intermediate terms in the computations of the correlator
increases very quickly, i.e., factorially. We have examined several
$r_0^{-6}$-order terms. We illustrate the computations with the
scalar scattering. The counter vertex operator at ${r_0^{-6}}$-order
is given by
%%%
 \bea
\pi V_{G,r_0^{-6}} \simeq
  %%%%%%%%%%%%%%%%%%%%
  &&= \fr{q}{r_0^2} \left(  \;-\fr{1}2\sqrt{h}\;h^{ij}
 \left(\pa_i X^u \pa_j X^u  X^n X^n
 -\pa_i X^m \pa_j X^m   \;X^n X^n\right)\right.
  \nn\\
 %%%%%%%%%%%%%%%%%%%%%%%%%%%%%%%%
 &&\left.-\fr{i\pa_i X^+}{2p^+}
   (\sqrt{h}\;h^{ij}-\ve^{ij})
  \left[X^n X^n\;S\pa_jS+\pa_jX^u X^n\; (S\g^{un}S))\right.\right.\nn\\
  &&\left.\left.-\pa_jX^m X^n\;(S\g^{mn}S) \right] \right. \nn\\
 %%%%%%%%%%%%%%%%%%%%%%%%%%%%%%%%%%
       &&\left.-\fr1{192}\sqrt{h}h^{ij}\fr{\pa_i X^+\pa_j X^+}{(p^+)^2}\;
       \{(S\g^{au} S)( S\g^{au} S) -  (S\g^{ab} S)( S\g^{ab} S)
       \}
 \right)
 \label{vgleading}
  \eea
  %%%
One of the correlators that we have considered is
 %%%
 \bea
&& \fr{q}{r_0^2}<V^{m_1}_s(x_1) V^{m_2}_s(x_2) V^{m_3}_s(x_3)
 V^{m_4}_s(x_4)\;\pa_i X^m \pa_j X^n \eta_{mn}X^lX^l>\nn\\
=&&\fr{q}{r_0^2}<X'^{m_1}X'^{m_2}X'^{m_3}X'^{m_4}\;\pa_i X^m \pa_j
X^m X^nX^n>
 \eea
 %%%
The other terms drop due to the dimensional regularization. It turns
out that the correlator vanishes
 %%%
 \bea
<X'^{m_1}X'^{m_2}X'^{m_3}X'^{m_4}\;\pa_i X^m \pa_j X^m X^nX^n>=0
 \eea
 %%%
 Therefore
 %%%
 \bea
&& \fr{q}{r_0^2}<V^{m_1}_s(x_1) V^{m_2}_s(x_2) V^{m_3}_s(x_3)
 V^{m_4}_s(x_4)\;\pa_i X^m \pa_j X^n \eta_{mn}X^lX^l>
=0
 \eea
 %%%
 As a matter of fact it is not too difficult to check that none of
 the ${r_0^{-6}}$-terms yields a finite result,
 %%%
 \bea
 <X'^{m_1}X'^{m_2}X'^{m_3}X'^{m_4}\;V_{G,r_0^{-6}}>=0
 \eea
 %%% at this order.
Another correlator that we have checked is  $<XXRR\;\pa X^u\pa X^u
X^nX^n>$. It also vanishes,
%%%
 \bea
 <XXRR\;\pa X^u\pa X^u X^nX^n>=0
 \eea
 %%%
In the higher order computations  it is often the high powers of
$\fr1{x_1}$ that are responsible for the null result since higher
order terms tend to come with higher powers of $\fr1{x_1}$. We
expect with reasonable confidence that all the higher order terms
will yield vanishing results because of this reason together with
the index structures.

In the introduction we have shown that the flat space is unable to
cancel the divergence due to a mismatch of a sign, which is
correctly produced by the action in the curved space. Together with
the results obtained in sec 3 we believe that it strongly supports
the notion of the engineering of the D-brane geometry by open string
loop effects. However, the fact that all the higher order
correlators checked so far vanish makes role-less all the terms in
$V_G$ that are more composite than those in quadratic order in
fields. It will be nice to see an example where that they do
contribute. It is likely that the non-contribution of the higher
order terms is a peculiar feature of the one-loop order. At higher
loop orders we expect that the presence and inter-correlation of the
cubic and more composite terms will be indispensable for the
cancelation of the divergence. It is one of the near-future
directions that we will pursue \cite{progress}.\footnote{For that
purpose it may be useful to attempt the corresponding computation in
the field theory context in an extension of the analysis that is
initiated in \cite{Park:2007ev}. } Another direction that we may
pursue is the computation of the open string analogue of the
anomalous dimension of N=4 SYM. It will be interesting to study
whether the full open string computation can lead to a resolution of
the three loop discrepancy.

Once we verify the conjecture with more examples of higher orders in
$r_0$ or/and $g$ one of the things that it establishes is the
picture that the open string, starting out in a flat space,
completes the theory toward the curved geometry. It will also imply
the relevance of the open string even in the final form of AdS/CFT
\cite{Park:1999xz}.\footnote{A related discussion can be found e.g.,
in \cite{Kawai:2007ek}.} The resulting non-linear sigma model then
will be analogous to the 1PI effective action in a quantum field
theory. The connection to AdS geometry and to AdS/CFT can be seen
through the effective field theory action, namely the DBI type
action, along the line of the following logic
\cite{Park:2001bm}\cite{Park:2008sg}. First apply an S-duality on
the DBI action making the coupling constant flip, $g\ra
g'=\fr{1}{g}$.  Then taking a $g\ra 0$ limit brings two things.
First, the D-brane geometry becomes an AdS space. Secondly, the
limit allows one to write down the solution of the equations of
motion of the DBI action in a particular form
\cite{Gibbons:2000hf}\cite{Sen:2000kd}, which, in turn, can be
interpreted as a closed string action.

Finally a few comments on the relation with the Fischler-Susskind
mechanism \cite{Fischler:1986ci,Fischler:1986tb} are in order. The
very idea of the role of the geometry in the divergence cancelation
is the same, in spirit, as that of the Fischler-Susskind mechanism.
There are a few differences as well. First of all, it is the set-up
of the computation, which in turn makes the interpretation of the
geometry very different. In \cite{Fischler:1986ci,Fischler:1986tb} (
a related discussion can be found in \cite{Das:1986dy}) the geometry
exists from the beginning as a fundamental object whereas in our
construction we are proposing that it should be a secondary
by-product of the flat-space loop effects. Secondly but perhaps more
importantly it is the relevance of the presence of a D-brane and the
transverse space. The analysis of
\cite{Fischler:1986ci,Fischler:1986tb,Das:1986dy} was carried out
for a closed string/space-time filling brane case. Therefore there
is no room for the transverse geometry. This is to be contrasted
with the present case where have a Dp-brane with $p<9$. Put in
another way, we do not expect the D9 brane to have non-trivial
geometry even in the higher order perturbation. Lastly, we note that
the geometry that results from the analysis of
\cite{Fischler:1986ci,Fischler:1986tb} is AdS/dS while in our case
it is (or is expected to be) the full Dp-brane geometry before the
S-duality that is mentioned in the previous paragraph.

\vspace{.5in}

\ni {\bf Acknowledgements:} I thank KIAS (Korea Institute for
Advanced Study) for its hospitality during my visit. Part of the
work was carried out during the stay.

\newpage

\renewcommand{\theequation}{A.\arabic{equation}}
 \setcounter{equation}{0}
  \section*{Appendix A: Derivation of geometry vertex operator
   }

\subsection{derivation of the action}

The GS action for a generic curved background was obtained by
several different groups \cite{Cvetic:1999zs}\cite{Sahakian:2004gy}
\cite{Mizoguchi:2002qy}. We use the action obtained by Sahakian
since his result includes fermionic quartic terms which is the
highest order, in the light-cone gauge, that can be present for a
certain class of configurations. We narrow down to the terms that is
relevant for the D3-brane geometry. The action that is zeroth order
in the fermionic coordinates
 is given, in our conventions, by
 %%%
 \bea
 S^{(0)}=-\int d^2\s\; \left[\fr12\sqrt{h}\;h^{ij}V_i^{\At} V_{\At j}
  -2\sqrt{h}\;h^{ij}V_i^+V_{j}^-\right]\label{s0}
 \eea
 %%%
where
 %%%
 \bea
 V_i^a\equiv \pa_i X^m e_m^a, \quad  V_i^\pm \equiv \pa_i X^m e_m^\pm,
 \eea
 %%%
with
 %%%
 \bea
 V_i^+=\fr{1}{{2}}(V_i^0+V_i^3)
 \eea
 %%%
Putting the quadratic fermionic terms and the quartic fermionic
terms together one gets
 %%%
 \bea
 && S^{(2)}+S^{(4)}\nn\\
 =&&\fr{1}{\pi}\int d^2\s\; -2i\sqrt{h}\;h^{ij}V_i^+
         (\th^1\pa_j\th^1+\th^2\pa_j\th^2) \nn\\
       && -\fr{i}2\sqrt{h}\;h^{ij}\pa_j X^{\Mt} w_{\Mt\Ct\Dt}
       V_i^+(\th^1\s^{\Ct\Dt}\th^1 + \th^2\s^{\Ct\Dt}\th^2 ) \nn\\
      &&-2i\;\ve^{ij}V_i^+
         (\th^2\pa_j\th^2-\th^1\pa_j\th^1) \nn\\
       && -\fr{i}2\;\ve^{ij}\pa_j X^{\Mt} w_{\Mt\Ct\Dt}
       V_i^+(\th^2\s^{\Ct\Dt}\th^2 - \th^1\s^{\Ct\Dt}\th^1 )\nn\\
  && +\fr{i}2 \ve^{ij}V_i^+V_j^{\Dt}{G^{-+\Ct}}_{\At\Bt}
         (\th^1\s^{\At\Bt}\th^2 + \th^2\s^{\At\Bt}\th^1
         )\;\eta_{\Ct\Dt}\nn\\
  %%%%%%%%%%%%%%%%%%%%%%%%%%%%%%%%%%%%
              &&+\sqrt{-h}\;h^{ij}V_i^{+}V_j^+ \{\nn\\
              &&     \fr{23}{576}(\th^1\th^2)^2\;G^{-+\At\Bt\Ct}{G^{-+}}_{\At\Bt\Ct}
                -\fr{1}{4608}(\th^t\s^{\At\Bt}\th^t)
                \;(\th^s\s_{\At\Bt}\th^s)
                G^{-+\Ct\Dt\Et}{G^{-+}}_{\Ct\Dt\Et}\nn\\
                &&+0
                  %-\fr{1}{384}(\thb\s^{-}\th)\;(\thb\s^{-\At\Bt}\th)
                  %\left[-\fr{1}{12}G^{-+\Ct\Dt\Et}G_{\Ct\Dt\Et\At\Bt}
                  %+{G^{-+\Ct}}_{\At\Dt}{G^{-+\Dt}}_{\Ct\Bt}\right]
                \nn\\
                &&-\fr{1}{768}(\th^t\s^{\At\Bt}\th^t)\;
                (\th^s\s^{\Ct\Dt}\th^s)
                \left[{G^{-+\Et}}_{\At\Bt}{G^{-+}}_{\Et\Ct\Dt}-\fr{1}{24}
                G_{\At\Bt\Et\Ft\Gt}{G^{\Et\Ft\Gt}}_{\Ct\Dt}\right]\nn\\
                &&+\fr{1}{128}(\th^t\s^{\At\Ct}\th^t)\;
                (\th^s{\s^{\Bt}}_{\Ct}\th^s)
                \left[{G^{-+}}_{\At\Dt\Et}{G^{-+\Dt\Et}}_{\Bt}-\fr{1}{72}
                G_{\At\Dt\Et\Ft\Gt}{G^{\Dt\Et\Ft\Gt}}_{\Bt}\right]\nn\\
         &&   -    \fr{1}{48} D_{\Ct}{G^{-+\Ct}}_{\At\Bt}\;\th^1\th^2\;
                         (\th^t\s^{\At\Bt}\th^t)    \nn\\
         &&         +\fr{5}{4}(\th^1\th^2)^2\;R^{-+-+}
                  +\fr{1}{96}(\th^t\s^{\At\Bt}\th^t)
                  (\th^s\s_{\At\Bt}\th^s) R^{-+-+}\nn\\
                  &&+\fr{1}{48}(\th^t\s^{\At\Ct}\th^t)\;
                (\th^s{\s^{\Bt}}_{\Ct}\th^s)
                 \left[{R^{-+}}_{\At\Bt}-\fr12
                  {R_{\At\Ct\Bt\Dt}}\eta^{\Ct\Dt}\right]\nn\\
                  &&+\fr{1}{192}(\th^t\s^{\At\Bt}\th^t)\;
                    (\th^s\s^{\Ct\Dt}\th^s)
                   \left[R_{\At\Ct\Bt\Dt}+\fr12
                  R_{\At\Bt\Ct\Dt}\right]\nn\\
                &&  \}
 \eea
 %%%
Due to the light-cone gauge constraint each fermionic coordinate
$\th^t$ has only eight non-zero components. Replacing the 16 by 16
gamma matrices $\s^{\At}$ by 8 by 8 matrices $\g^{\At}$ one gets,
after
 %%%
 \bea
 X^+=\fr{p^0+p^3}{2},\quad\th\th \ra \fr{1}{2p^+}SS,
 \eea
%%%
%%%
 \bea
 && S^{(2)}+S^{(4)}\nn\\
 =&&\fr{1}{\pi}\int d^2\s\; \fr1{2p^+}\{-2i\sqrt{h}\;h^{ij}V_i^+
         (S^t\pa_jS^t) \nn\\
       && -\fr{i}2\sqrt{h}\;h^{ij}\pa_j X^{\Mt} w_{\Mt\Ct\Dt}
       V_i^+(S^t\s^{\Ct\Dt}S^t) \nn\\
      &&-2i\;\ve^{ij}V_i^+
         (S^2\pa_jS^2-S^1\pa_jS^1) \nn\\
       && -\fr{i}2\;\ve^{ij}\pa_j X^{\Mt} w_{\Mt\Ct\Dt}
       V_i^+(S^2\s^{\Ct\Dt}S^2 - S^1\s^{\Ct\Dt}S^1 )\nn\\
  && +\fr{i}2 \ve^{ij}V_i^+V_j^{\Dt}{G^{-+\Ct}}_{\At\Bt}
         (S^2\s^{\At\Bt}S^1 + S^1\s^{\At\Bt}S^2
         )\;\eta_{\Ct\Dt}\nn\\
         &&\}\nn\\
  %%%%%%%%%%%%%%%%%%%%%%%%%%%%%%%%%%%%
              &&+\fr1{4(p^+)^2}\sqrt{-h}\;h^{ij}V_i^{+}V_j^+ \{\nn\\
              &&     \fr{23}{576}(S^1S^2)^2\;G^{-+\At\Bt\Ct}{G^{-+}}_{\At\Bt\Ct}
                -\fr{1}{4608}(S^t\s^{\At\Bt}S^t)
                \;(S^s\s_{\At\Bt}S^s)
                G^{-+\Ct\Dt\Et}{G^{-+}}_{\Ct\Dt\Et}\nn\\
                &&+0
                  %-\fr{1}{384}(Sb\s^{-}S)\;(Sb\s^{-\At\Bt}S)
                  %\left[-\fr{1}{12}G^{-+\Ct\Dt\Et}G_{\Ct\Dt\Et\At\Bt}
                  %+{G^{-+\Ct}}_{\At\Dt}{G^{-+\Dt}}_{\Ct\Bt}\right]
                \nn\\
                &&-\fr{1}{768}(S^t\s^{\At\Bt}S^t)\;
                (S^s\s^{\Ct\Dt}S^s)
                \left[{G^{-+\Et}}_{\At\Bt}{G^{-+}}_{\Et\Ct\Dt}-\fr{1}{24}
                G_{\At\Bt\Et\Ft\Gt}{G^{\Et\Ft\Gt}}_{\Ct\Dt}\right]\nn\\
                &&+\fr{1}{128}(S^t\s^{\At\Ct}S^t)\;
                (S^s{\s^{\Bt}}_{\Ct}S^s)
                \left[{G^{-+}}_{\At\Dt\Et}{G^{-+\Dt\Et}}_{\Bt}-\fr{1}{72}
                G_{\At\Dt\Et\Ft\Gt}{G^{\Dt\Et\Ft\Gt}}_{\Bt}\right]\nn\\
         &&   -    \fr{1}{48} D_{\Ct}{G^{-+\Ct}}_{\At\Bt}\;S^1S^2\;
                         (S^t\s^{\At\Bt}S^t)    \nn\\
         &&         +\fr{5}{4}(S^1S^2)^2\;R^{-+-+}
                  +\fr{1}{96}(S^t\s^{\At\Bt}S^t)
                  (S^s\s_{\At\Bt}S^s) R^{-+-+}\nn\\
                  &&+\fr{1}{48}(S^t\s^{\At\Ct}S^t)\;
                (S^s{\s^{\Bt}}_{\Ct}S^s)
                 \left[{R^{-+}}_{\At\Bt}-\fr12
                  {R_{\At\Ct\Bt\Dt}}\eta^{\Ct\Dt}\right]\nn\\
                  &&+\fr{1}{192}(S^t\s^{\At\Bt}S^t)\;
                    (S^s\s^{\Ct\Dt}S^s)
                   \left[R_{\At\Ct\Bt\Dt}+\fr12
                  R_{\At\Bt\Ct\Dt}\right]\nn\\
                &&  \}
 \eea
 %%%
The IIB super-gravity solution for the D3 brane configuration is
given by
 %%%
 \bea
ds^2 &=& H^{-1/2}(dx^\m)^2+H^{1/2}(dx^m)^2 \nn\\
 G_{\bar{0}\bar{1}\bar{2}\bar{3}c}&=&-\fr{X^c}{r}H^{-5/4}H'\nn\\
  G_{abcde} &=& \fr{H^{-5/4}H'}{r}\ve_{abcdef}X^f
 \eea
 %%%
The bar indicates that the indices are flat. Substituting the
explicit forms of the connection, the five form and the Riemann
tensor into the total action one gets
 %%%
 \bea
 && { S}^{(0)}+{ S}^{(2)}+{S }^{(4)} \nn\\
 =&&\fr1{\pi}\int d^2\s -\left[\fr12\sqrt{h}\;h^{ij}V_i^{\At} V_{\At j}
  -2\sqrt{h}\;h^{ij}V_i^+V_{j}^-\right]\nn\\
 %%%%%%%%%%%%%
 && + \fr1{2p^+}\{-2i\sqrt{h}\;h^{ij}V_i^+
         (S^t\pa_jS^t) -\;2i\ve^{ij}V_i^+
         (    S^2\pa_jS^2-S^1\pa_jS^1) \nn\\
 &&+\fr{i}4 \sqrt{h}h^{ij}V_i^+ H^{-3/2}\fr{H'}{r} \pa_jX^u
        X^mS^t\g^{um}S^t\nn\\
  &&    -\fr{i}4 \sqrt{h}h^{ij}V_i^+ H^{-1}\fr{H'}{r}
       \pa_jX^m X^nS^t\g^{mn}S^t\nn\\
     &&+\fr{i}4V_i^+\ve^{ij}  H^{-3/2}\fr{H'}{r} \pa_jX^u
        X^m (S^2\g^{um}S^2-S^1\g^{um}S^1)\nn\\
      && -\fr{i}4 V_i^+ \ve^{ij}H^{-1}\fr{H'}{r}
       \pa_jX^m X^n(S^2\g^{mn}S^2-S^1\g^{mn}S^1)\nn\\
  && +i\ve^{ij}V_i^+ H^{-5/4}\fr{H'}{r}X^b
             (V_j^b S^2\g^{12}S^1+V_j^1S^2\g^{2b}S^1
             -V_j^2S^2\g^{1b}S^1)\nn\\
       &&\}\nn\\
   %%%%%%%%%%%%%%%%%%%%%%%%%%%
       &&+\fr1{4(p^+)^2}\sqrt{h}h^{ij}V_i^+V_j^+\;\;\{\nn\\
  &&  \fr{23}{24}(S^1S^2)^2H^{-5/2}(H')^2
  -\fr{1}{192}(S^t\g^{\At\Bt} S^t)( S^{t'}\g_{\At\Bt} S^{t'})
               H^{-5/2}(H')^2 \nn\\
 &&-\fr{1}{768}H^{-5/2}(H')^2
   \left[16(S^t\g^{12} S^t)( S^{t'}\g^{12} S^{t'})
   +16(S^t\g^{au} S^t)( S^{t'}\g^{cu} S^{t'})\fr{X^aX^c}{r^2}\right.\nn\\
   &&\left. \hspace{1.3in}-\fr{1}{24}(S^t\g^{ab} S^t)( S^{t'}\g^{cd} S^{t'})
    \fr{X^hX^{h'}}{r^2}\ve_{abefgh}\ve_{cdefgh'}
   \right]\nn\\
 &&+\fr{1}{16}H^{-5/2}(H')^2
   \left[(S^t\g^{a\Ct} S^t)( S^{t'}{\g^{b}}_{\Ct} S^{t'})\fr{X^aX^b}{r^2}
   +(S^t\g^{au} S^t)( S^{t'}\g^{au} S^{t'})\right.\nn\\
   &&\left.
  \hspace{1.3in} -\fr{1}{576}(S^t\g^{a\Ct} S^t)( S^{t'}{\g^{b}}_{\Ct} S^{t'})
    \fr{X^hX^{h'}}{r^2}\ve_{adefgh}\ve_{bdefgh'}
 \right]
   \nn\\
    &&+\fr{1}{24}H^{-5/2}(H')^2\; S^1S^2(S^t\g^{12} S^t)
    +\fr{5}{16}H^{-5/2}(H')^2\; S^1S^2\nn\\
 &&+\fr{1}{384}H^{-5/2}(H')^2\; (S^t\g^{\At\Bt} S^t)(S^{t'}\g_{\At\Bt} S^{t'})
 \nn\\
 &&+\fr{3}{1536}H^{-5/2}(H')^2(S^t\g^{\a\Ct} S^t)(S^{t'}\g_{\a\Ct}
 S^{t'})\nn\\
 &&-\fr{1}{96}(4g_2X^aX^b+4g_3\d_{ab})
 (S^t\g^{a\Ct} S^t)(S^{t'}{\g^{b}}_{\Ct} S^{t'})\nn\\
 &&-\fr{1}{384}(4g_2r^2+24g_3)\; (S^t\g^{\a\Ct} S^t)(S^{t'}\g_{\a\Ct}
 S^{t'})\nn\\
&& -\fr{1}{96}(4h_1X^aX^b+h_1r^2\d_{ab}+5h_2\d_{ab})
 (S^t\g^{a\Ct} S^t)(S^{t'}{\g^{b}}_{\Ct} S^{t'})\nn\\
 &&+\fr{1}{96}g_1(S^t\g^{\a\b} S^t)(S^{t'}\g_{\a\b} S^{t'})
 +\fr{1}{48}(g_2X^aX^b+g_3\d_{ab})\; (S^t\g^{\a a}
 S^t)(S^{t'}{\g_{\a}}^b
 S^{t'})\nn\\
 &&+\fr{1}{96}\left[2h_1X^aX^b(S^t\g^{ae} S^t)(S^{t'}\g_{be} S^{t'})
 +h_2(S^t\g^{ab} S^t)(S^{t'}\g_{ab} S^{t'})\right]
 \nn\\
 &&\} \label{startingaction}
 \eea
 %%%
When we compute the amplitude below we will use the dimensional
regularization. Considering that the scattering states contain only
the $S^1$ coordinate but not $S^2$ we can drop the terms in
(\ref{startingaction}) that have an $S^2$ factor. Defining
 %%%
 \bea
 S\equiv S^1
 \eea
 %%%
 and setting $V_j^-=0$, one gets after some algebra
%%%
 \bea
 && { S}^{(0)}+{ S}^{(2)}+{S }^{(4)} \nn\\
 =&&\fr1{\pi}\int d^2\s \;-\!\!\fr{1}2\sqrt{h}\;h^{ij}
 \left(\pa_i X^u \pa_j X^v \eta_{uv} H^{-1/2}+
 \pa_i X^m \pa_j X^n \eta_{mn} H^{1/2}\right)
  \nn\\
 %%%%%%%%%%%%%%%%%%%%%%%%%%%%%%%%
 && +\fr1{2p^+} \{ \nn\\
  && -2i(\sqrt{h}\;h^{ij}-\ve^{ij})\pa_i X^+ H^{-1/4}
         (S\pa_jS)  \nn\\
 &&+\fr{i}4 (\sqrt{h}h^{ij}-\ve^{ij})\pa_i X^+ H^{-7/4}\fr{H'}{r} \pa_jX^u
        X^m\; (S\g^{um}S)\nn\\
  &&    -\fr{i}4 (\sqrt{h}h^{ij}-\ve^{ij})\pa_i X^+ H^{-5/4}\fr{H'}{r}
       \pa_jX^m X^n\;(S\g^{mn}S)\nn\\
        && \quad\quad\} \nn\\
 %%%%%%%%%%%%%%%%%%%%%%%%%%%%%%%%%%
       &&+\fr1{4(p^+)^2}\sqrt{h}h^{ij}\pa_i X^+\pa_j X^+\;H^{-1/2}\{\nn\\
  &&-\fr{17}{1536}\k_1(S\g^{uv} S)( S\g^{uv} S)+\left[
  \fr{43}{768}\k_1+\fr1{192}\k_2
  \right](S\g^{au} S)( S\g^{au} S) \nn\\
  &&-\left[
  \fr1{192}\k_2 +\fr1{128}\k_1\right] (S\g^{ab} S)( S\g^{ab} S)\nn\\
  &&+X^aX^b \fr{1}{r^2}\left[\fr{31}{768}\k_1-\fr1{32}\k_2
  \right](S\g^{au} S)( S\g^{bu} S)\nn\\
  &&+X^aX^b \fr{1}{r^2}\left[
  +\fr1{32}\k_2+\fr{29}{384}\k_1\right](S\g^{ac} S)( S\g^{bc} S) \nn\\
 &&\} \label{startingaction5}
 \eea
 %%%
where
 %%%
 \bea
 \k_1=H^{-5/2}(H')^2,\quad\quad \k_2=H^{-3/2}H'\fr1r
 \eea
 %%%

\newpage
%%%%%%%%%%%%%%%%%%%%%%%%%%%%%%%%%%%%

\end{document}